\documentclass[pre, superscriptaddress]{revtex4}
\usepackage{amsmath}                   
\usepackage{amsfonts}                  
\usepackage{epsfig}                    
\usepackage{latexsym}

\DeclareMathOperator{\Adj}{Adj}
\DeclareMathOperator{\Trace}{Trace}

\begin{document}
\title{Random copolymer adsorption: Morita approximation compared to exact
numerical simulations}

\author{Alexey Polotsky}
\email{alexey.polotsky@gmail.com}
\affiliation{Institute of Macromolecular Compounds, Russian Academy of Sciences 31 Bolshoy pr., 199004 St.-Petersburg, Russia}

\author{Andreas Degenhard}
\email{adegenha@Physik.Uni-Bielefeld.de}
\affiliation{Fakult\"at f\"ur Physik, Universit\"at Bielefeld,
Universit\"atsstra{\ss}e 25, D-33615 Bielefeld, Germany}

\author{Friederike Schmid}
\email{schmid@Physik.Uni-Bielefeld.de}
\affiliation{Institut f\"ur Physik, Universit\"at Mainz,
Staudingerweg 7, D-55099 Mainz, Germany}

\begin{abstract}
We study the adsorption of ideal random lattice copolymers with
correlations in the sequences on homogeneous substrates with
two different methods: An analytical solution of the problem
based on the constrained annealed approximation introduced by Morita
in 1964 and the generating functional (GF) technique, and direct
numerical simulations of lattice chains averaged over many
realizations of random sequences. Both methods allow to calculate the
free energy and different conformational characteristics of the
adsorbed chain.  The comparison of the results for random copolymers
with different degree of correlations and different types of nonadsorbing
monomers (neutral or repelling from the surface) shows not only qualitative
but a very good quantitative
agreement, especially in the cases of Bernoullian and
quasi-alternating random sequences.

\end{abstract}

\maketitle

\section{Introduction}
Random copolymers (RC) attract much attention and have been studied intensively
in the last two decades. This is motivated, on the one hand, by the nontrivial
properties of individual RCs and their behaviour in solutions and melts, and,
on the other hand, by the biological relevance of this class of molecules
(protein \cite{Pande:2000} and RNA \cite{Bundschuh:2002} folding,
hybridisation of heterogeneous DNA \cite{Garel:2005}, because real proteins
and nucleic acids represent, with no doubt, irregular copolymers.

Among the many open questions involving RCs, the problem of RC
adsorption at different interfaces is taking a particularly
prominent place, because of the technological relevance of such
systems, their role as biomimetic model systems for molecular
recognition \cite{Chakraborty:2001} etc.), and the basic challenges
they offer for theoretical physicists and applied mathematicians.
Several types of interfaces have been considered, interfaces that
are penetrable for polymer such as liquid/liquid interfaces or lipid
membranes, and impenetrable solid substrates, either with or without
chemical or geometrical homogeneities \cite{Sommer:1995,Trovato:1999,Chen:2000,
Denesyuk:2000, Polotsky:2004a, Polotsky:2004b,Orlandini:2002,
Iliev:2004,Alvarez:2007}. In the present paper we focus on the problem 
of adsorption of a single ideal RC chain with correlations in the 
monomer sequence on an impenetrable planar surface using a lattice 
model.

Solving problems that involve RCs, one faces the necessity to carry out two
operations: Firstly, one needs to sum over the statistical weights of polymer
conformations, i.e. to calculate the system's partition function $Z$ and
the free energy $F$ (the logarithm of the partition function, $F=-k_B T \log Z$),
and secondly, one needs to average \emph{the free energy} with respect
to the disorder in the RC monomer sequence. Averaging ``$\log Z$'' over the
sequence disorder turns out to be a very difficult task which can seldom be
carried out rigorously. It would be much easier to average the partition function
prior to taking its logarithm, but this corresponds to the physically different
situation of \emph{annealed disorder} (in our system the monomer sequence
is \emph{quenched}), which approximates the quenched situation only very poorly.
One common approach consists in using the replica trick which then can be
combined with some variational scheme ({\em e.g.}, the reference system approach
suggested by Chen \cite{Chen:2000, Denesyuk:2000, Polotsky:2004a, Polotsky:2004b},
or the Gaussian variational approach \cite{Trovato:1999}).

In the present work, we exploit another approximation to resolve the difficulties
with the quenched average - the constrained annealing suggested by T. Morita in
1964~\cite{Morita:1964}. In this approximation scheme, one averages over annealed
disorder ({\em i.e.} directly averages the partition function), but imposes the
constraint that the first, the second, etc., moments of the monomer distribution
in the RC sequence keep their correct values. The ideas of the method are developed
in the original paper of Morita~\cite{Morita:1964}; a more detailed and formalized
description can be found in the paper of K\"uhn~\cite{Kuehn:1996} and in a recent
review of Soteros and Whittington~\cite{Soteros:2004}.

For the problem of RC adsorption on the solid interface and RC localization at the
liquid-liquid interface, the Morita approximation has been successfully applied by
Whittington and Orlandini \emph{et al.} \cite{Orlandini:2002,Iliev:2004}.
In these works, the authors considered simplified polymer models - fully directed
random walks in 2D space (Dyck and Motzkin paths). RC localization at penetrable
and impenetrable interfaces \cite{Orlandini:2002} (with both chain ends fixed at
the interface) as well as the adsorption-desorption transition upon pulling one chain
end away from the interface by an external force~\cite{Iliev:2004} were studied.
The Morita approximation was first order, {\em i.e.}, only a constraint on the
first moment of the monomer distribution was imposed. (Note that the annealed
approximation can formally be considered as a zero order Morita approximation.)

Higher order Morita approximations were applied to the RC adsorption
problem by Alvarez \emph{et al.}~\cite{Alvarez:2007}, where polymer
trajectories were again modeled by Dyck and Motzkin paths. The
authors obtained lower bounds limiting the quenched free energy and
demonstrated how these bounds improve as the order of the Morita
approximation increases. In contrast, the location of the phase
boundary is not more accurate in the Morita approximation than in
the annealed approximation, as was rigorously proved by Caravenna
and Giacomin~\cite{Caravenna:2005}

The simplified polymer models (Dyck and Motzkin paths) considered in the
previous works~\cite{Orlandini:2002, Iliev:2004, Alvarez:2007} have their
doubtless advantages of being exactly solvable and at the same time permitting
to obtain a physically clear and tractable picture of the studied phenomena.
However, real polymers ``live'' in the 3-dimensional space and are not necessarily directed.
Furthermore, the RCs in Refs.~\onlinecite{Orlandini:2002, Iliev:2004, Alvarez:2007}
were of Bernoullian type, {\em i.e.}, uncorrelated. In the present paper we study
-- although somewhat simplified -- a 3-dimensional situation and a wider class
of RCs with correlations in the monomer sequence, where Bernoullian RCs represent
one special case.

The calculation of the partition function via summation over polymer
conformations is carried out using the generating function (GF)
approach (or grand canonical approach) developed for polymeric
systems by Lifson~\cite{Lifson:1964}. This very general method has
been used to study conformational transitions in polypeptides (coil
-- $\alpha$-helix~\cite{Poland:1965} and  coil --
$\beta$-structure~\cite{Birshtein:1971}) and DNA (coil -- double
helix) \cite{Litan:1965, Poland:1966}. Birshtein was the first to
apply it to the problem of single polymer
adsorption~\cite{Birshtein:1979, Birshtein:1983}. She developed a
theory of the adsorption transition based on the GF formalism and
demonstrated its power and generality with a number of different
examples. The approach allowed to reproduce very elegantly a number
of previously known results, and, in the same easy manner, to treat
the more complicated case when adsorption is coupled with a
helix-coil transition in the chain. In her work
Birshtein~\cite{Birshtein:1979, Birshtein:1983} focused on the
adsorption-desorption transition point and the order of the
transition. Later, the results of her theory were used to analyze
the adsorption-desorption transition for regular $\mathrm{\left(A_m
B_n\right)_x}$-multiblock copolymers \cite{Zhulina:1981}. Brun
\cite{Brun:1999} has proposed an approximate generalization of the
equation for the transition point in
Ref.~\onlinecite{Birshtein:1979} for correlated annealed RCs (see
section \ref{subs:annealed}).

The method of GF is highly general and its success is due to the possibility to
calculate the GF for \emph{elements} of the chain conformations. Despite the fact
that it is more than 30 years old, it is still often used up to date. One of the
most prominent examples is the ongoing discussion about the order of the
DNA denaturation transition \cite{Kafri:2000, Kafri:2002, Garel:2001, Richard:2004},
which is carried on - for the most part - in the framework of this approach.
We also note that in its ``continuous'' version, the method of GF is equivalent
to the propagator formalism in Fourier-Laplace space -- see, for instance,
Muthukumar et al.~\cite{Muthukumar:1996, Carri:1999}.

In the present work, analytical calculations based on the Morita approximation
and the GF approach (though the final equations can only be solved numerically,
the disorder average and the summation over chain conformations are carried out
analytically) are supplemented with numerical simulations for tethered RC chains.
In the simulations, a sample of several (sufficiently many) RC sequences
are drawn randomly from the desired distribution. For each chain, the propagators
are calculated exactly (recursively). This allows to evaluate the free energy and
different observables directly and carry out an ``exact'' \emph{quenched} average
over the sequence disorder. The comparison of the results obtained with the two
different methods allows us to make a conclusion about the accuracy and applicability of
the Morita approximation.

The rest of the paper is organized as follows.  Section~\ref{sec:model} 
defines the model and introduces the annealed and the Morita approximation. 
Although the annealed approximation is deficient in many respect, it is 
convenient to introduce the GF approach in this framework. This is done 
in Section~\ref{sec:gf}. An additional motivation for treating the annealed 
case in detail is that there is a class of so called two-state or annealed 
copolymers, which are adequately described by the annealed approximation 
(see~Ref.~[\onlinecite{Yoshinaga:2008}] and references therein). The 
generalisation to the Morita approximation in Section~\ref{sec:gf} is 
quite simple and transparent. In Section~\ref{sec:numerics} our numerical 
approach is introduced. The results obtained using both methods are compared 
and discussed in Section~\ref{sec:results}.  Finally, 
Section~\ref{sec:summary} presents the summary and final conclusions.

\section{Model and method} \label{sec:model}
\subsection{Definition of the model} \label{sec:model_def}
\begin{figure}[t]
  \begin{center}
  \includegraphics[width=8cm]{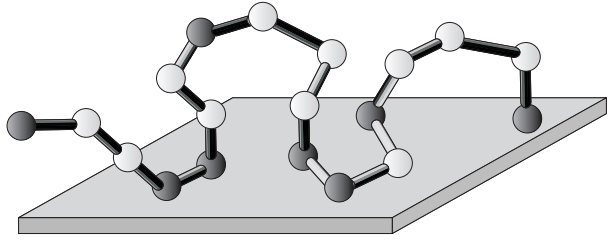}
  \end{center}
  \caption{Random copolymer near adsorbing plane.}
  \label{fig:RC}
\end{figure}

Consider a RC chain consisting of $N$ monomer units near an impenetrable  surface,
Fig.~\ref{fig:RC}. A lattice model of the polymer is employed, i.e.  each chain
conformation is represented as a walk on the lattice. Monomer units in the RC chain
can either be of type A or of type B, and their sequence is taken to be random.
We assume that there are no excluded-volume interactions between monomers ({\em i.e.},
we consider the case of ideal chains), but there is a short-range interaction between
monomers and the surface. The Hamiltonian of the system can be written as
\begin{equation} \label{eq:H}
   H = \sum_{i=1}^{N} \Delta_i \cdot \left[ \chi_i \varepsilon_A  +
       (1-\chi_i)\varepsilon_B \right] .
\end{equation}
Here $\chi_i$ indicates the type of the monomer $i$:
\begin{equation} \label{eq:chi_def}
   \chi_i=\left\{
      \begin{array}{ll}
         1 & \mbox{for A-monomer } \left(\mbox{giving } [...]=\varepsilon_A \right) \\
         0 &  \mbox{for B-monomer } \left(\mbox{giving } [...]=\varepsilon_B \right)
      \end{array}
   \right.
\end{equation}
and $\Delta_i$ is a conformation-dependent parameter, namely, $\Delta_i=1$ if $i$-th
monomer occupies a site adjacent to the adsorbing surface and 0 otherwise.
Any contact of an A monomer with the surface leads to an energy contribution
$\varepsilon_A$, and likewise, a contact of a B monomer contributes $\varepsilon_B$.

The monomer sequence $\chi=\{\chi_1, \chi_2 , ... , \chi_N\}$ is random and will be
represented as a first order Markov chain. It is determined by the probabilities
to find A and B monomers in the sequence
\begin{equation} \label{eq:f:def}
   P(\chi_i = A)=f_A \, \mbox{ and } \, P(\chi_i = B)=f_B=1-f_A
\end{equation}
and by the probabilities that the monomer of the type $i$ is followed by the monomer of
the type $j$, $P(\chi_m=j|\chi_{m-1}=i)=p_{ij}$
\begin{equation} \label{eq:P:def}
   \mathbf{P}=
   \left(
      \begin{array}{cc}
         p_{AA} & p_{AB} \\
         p_{BA} & p_{BB}
      \end{array}
   \right) =
   \left(
      \begin{array}{cc}
         p_{AA} & 1-p_{AA} \\
         1-p_{BB} & p_{BB}
      \end{array}
   \right) .
\end{equation}
The transition probabilities are normalized : $\sum_j p_{ij}=1$.
We also assume that the Markov chain is reversible, i.e.
$f_i p_{ij} = f_j p_{ji}$. Hence,the probability of the sequence $\chi=\{\chi_1, \chi_2 , ... , \chi_N\}$ to appear is given by the product
\begin{equation} \label{eq:seqprob}
P(\chi) \equiv P(\chi_1, \chi_2, \ldots, \chi_n) =
f_{\chi_1}\cdot p_{\chi_1 \chi_2}\cdot p_{\chi_2 \chi_3} \cdot
  \ldots \cdot p_{\chi_{n-1}, \, \chi_n}.
\end{equation}

It is convenient to introduce the cluster parameter $c$:
\begin{equation}
    c := 1 - p_{AB} - p_{BA}=p_{AA} + p_{BB} - 1,
\end{equation}
which characterizes the correlations in the sequence: $c>0$ means
that there is a tendency in the sequence for grouping similar
monomers into clusters, $c<0$  favors the alternating sequence of
A's and B's, $c=0$ corresponds to uncorrelated (Bernoullian)
sequences. The cases $c=-1$ is also special, since it describes
perfectly ordered alternating block copolymers ($...ABABAB...$).

Because of the normalization and reversibility conditions, it is sufficient to prescribe
any two of these parameters to completely determine the first order Markov chain, for
example, $f_A$ and $c$. Then
\begin{equation}
  \begin{array}{ll}
    p_{AA}=f_A + c(1-f_A) \, , & p_{AB}=(1-f_A)(1-c) \\
    p_{BA}=f_A(1-c) \, ,       & p_{BB}=1-f_A + c f_A .
  \end{array}
\end{equation}

Finally, it should be noted that the range of possible values for $f_A$ depends on the
value of $c$: for $c\ge 0$ $f_A$ may take any value between 0 and 1 whereas for
$c<0$ there is a constraint
\begin{equation} \label{eq:fa:range}
    -\frac{c}{1-c} < f_A < \frac{1}{1-c} \, .
\end{equation}

In the actual calculations of this paper, we consider two cases:
\begin{itemize}
    \item A monomer is attracted by the surface ($\varepsilon_A = -1$),
              B monomer is neutral ($\varepsilon_B = 0$),
          denoted as SN (sticky-neutral) case
    \item A monomer is attracted by the surface ($\varepsilon_A = -1$),
              B monomer is repelled from the surface ($\varepsilon_B = 1$),
          denoted as SR (sticky-repulsive) case
\end{itemize}

It should be emphasized that the terms ``neutral'' or ``repulsive''
are applied to monomer-surface interaction whereas a polymer chain
composed of either neutral or repulsive monomers will effectively
repel from the surface.

\subsection{Disorder average and Morita approximation}
 \label{sec:morita}
Having introduced the model, let us now write the system's partition function for a given
realization of the monomer sequence
\begin{equation} \label{eq:Z:def}
Z_N(\beta|\chi) = \sum_{\omega} \exp\left\{-\beta \sum_{i=1}^{N}
  \Delta_i \cdot \left[ \chi_i \varepsilon_A  + (1-\chi_i)\varepsilon_B \right] \right\},
\end{equation}
where $\beta$ is the inverse temperature and $\sum_\omega$ denotes the sum
over all chain conformations $\omega$.

Rather than dealing with particular realizations of a RC sequence, we study the whole
ensemble of RC chains with the desired statistics. The randomness in the RC sequence is
\textit{quenched}, which means that after synthesis of a RC chain, its monomer sequence
remains unchanged. To obtain the quenched free energy, the logarithm of the partition
function should be averaged over all possible sequence realizations:
\begin{equation} \label{eq:Zq:def}
\beta F_q = - \left\langle \log Z_N(\beta|\chi) \right\rangle
\end{equation}
where the angular brackets $\left\langle \ldots \right\rangle$ denote averaging over
sequence randomness
\begin{equation}
\left\langle A \right\rangle := \sum_\chi P(\chi) A(\chi).
\end{equation}
Direct averaging of the $\log Z_N $ is, in general, a rather difficult problem, so one
can use, for example, the replica trick and work formally with an $m$ times replicated
system~\cite{Soteros:2004, Chen:2000, Stepanow:2002, Polotsky:2004a, Polotsky:2004b}.

Averaging of the partition function prior to taking the logarithm corresponds to the
\emph{annealed} type of disorder. In this case, the average is evaluated according to
\begin{equation} \label{eq:Fa}
\beta F_a = - \log \left\langle Z_N(\beta|\chi) \right\rangle,
\end{equation}
where
\begin{equation} \label{eq:Za:def}
  \left\langle Z_N(\beta|\chi) \right\rangle = \sum_{\chi} P(\chi) \sum_{\omega} \exp \left\{
  -\beta \sum_{i=1}^{N} \Delta_i \cdot \left[ \chi_i \varepsilon_A  + (1-\chi_i)\varepsilon_B
  \right] \right\}.
\end{equation}
Physically, annealed disorder means that the type of any monomer (and its affinity to the surface, in our case) may change while the system attains its equilibrium state~\cite{Yoshinaga:2008}. The annealed approximation can roughly be considered as the ``zero order'' approach to the problem of carrying out a quenched average.
It is substantially easier to perform but it often fails to give good values for the free energy and other observables. The annealed approximation does not guarantee that even the lower moments of the distribution of monomers remain correct in the final result.

To approach this problem, Morita~\cite{Morita:1964} has suggested to use a
\emph{partial annealing} procedure where the moments of the monomer distribution are
constrained to keep their correct values. For example, in our case, the constraint
on the first moment reads
\begin{equation} \label{eq:constr1}
\frac{1}{N} \sum_{i=1}^{N} \chi_i = f_A,
\end{equation}
the second moment of the distribution is constrained according to
\begin{equation} \label{eq:constr2}
\frac{1}{N-1} \sum_{i=2}^{N} \chi_{i-1} \chi_i = f_{AA},
\end{equation}
where $f_{AA}=f_A \cdot p_{AA}$ is the AA dyad probability, and so on.
For the averaging of the partition function, this means that we introduce the
constraints (\ref{eq:constr1})-(\ref{eq:constr2}) into Eq. (\ref{eq:Za:def})
via Lagrange multipliers $\lambda$ and $\kappa$:
\begin{equation} \label{eq:ZM2:def}
\begin{split}
\left\langle Z_N(\beta|\chi) \right\rangle_M =
& \sum_{\chi} P(\chi) \sum_{\omega} \exp \left\{-\beta \sum_{i=1}^{n} \Delta_i
\cdot \left[ \chi_i \varepsilon_A  + (1-\chi_i)\varepsilon_B \right] \right\} \\
\times & \, \exp \left\{  \lambda \left( \sum_{i=1}^{N} \chi_i - N f_A \right)
+ \kappa \left( \sum_{i=2}^{N} \chi_{i-1} \chi_i - (N-1) f_{AA} \right) \right\} .
\end{split}
\end{equation}
and request
\begin{equation} \label{eq:MConstr:deriv}
\begin{split}
& \frac{\partial \log \left\langle Z_N(\beta|\chi) \right\rangle_M}{\partial \lambda} =
\left\langle \sum_{i=1}^{N} \chi_i   \right\rangle - N f_A  = 0, \\
& \frac{\partial \log \left\langle Z_N(\beta|\chi) \right\rangle_M}{\partial \kappa} =
\left\langle \sum_{i=2}^{N} \chi_{i-1} \chi_i \right\rangle - (N-1) f_{AA}
= 0 \,.
\end{split}
\end{equation}
Here and below the notation $\left\langle \ldots \right\rangle_M$
for the disorder average subjected to the Morita conditions is used.
After exchanging the order of conformational and disorder average
(i.e. the order of summation $\sum_\chi \sum_\omega = \sum_\omega
\sum_\chi$), the partition function reduces to the sum of the
pre-averaged statistical (Boltzmann) weights over polymer
conformations. In the present work this sum is calculated using the
GF approach (or grand canonical approach) developed for linear
polymers by Lifson~\cite{Lifson:1964}.

\subsection{Sum over conformation and GF approach}
 \label{sec:gf}
The main object (concept) of the GF approach is the GF
\begin{equation} \label{eq:GF:def}
\Gamma(x)=\sum_{N=1}^\infty Z_N  x^N \, ,
\end{equation}
where $Z_N$ is the partition function of a polymer chain with $N$
units. The GF can be considered as the grand canonical partition
function; in this case, $x\equiv\exp(\mu/k_BT)$, where $\mu$ is the
chemical potential of a monomer unit and $x$ is called monomer
activity. Once the GF~(\ref{eq:GF:def}) is known, the partition
function $Z_N$ can be formally determined as the coefficients of the
expansion of $\Gamma(x)$ in powers of $x$. In the long chain limit,
$N\gg 1$, the asymptotic expression for $Z_N$ is even simpler since
it is dominated by the \emph{smallest singularity} $x_c$ of
$\Gamma(x)$: $Z_N \simeq x_c^{-N}$.

The GF approach is particularly efficient in the cases where monomer
unit may exist in one of several states (for example, double helical
or coil in DNA~\cite{Litan:1965, Poland:1966}, $\alpha$-helical and
coil \cite{Poland:1965} or $\beta$-structure and
coil~\cite{Birshtein:1971} in polypeptides) and the chain
conformation can be represented as an alternation (not necessary
regular) of sequences of different types.

To study RC adsorption, we generalize the work of Birshtein, who was
the first to apply the GF formalism to the problem of homopolymer
adsorption \cite{Birshtein:1979, Birshtein:1983}. For clarity, we
first illustrate our approach for the case of annealed RCs.

For an adsorbed chain, each adsorbed conformation can be represented
as a sequence of adsorbed and desorbed groups of monomers. The
adsorbed sequences are called \emph{trains}, among the desorbed we
distinguish between \emph{tails} at the ends of the chain and
\emph{loops} separating adsorbed sequences. It is clear that an
adsorbed chain can have at most two tails, $n_T \leq 2$, whereas the
number of loops is less than the number of trains by unity, i.e. if
$n_L=n_S-1$. We note that different tails, loops, and trains do not
interact with each other. The GF of the partition function, averaged
over (annealed) sequence disorder, $\Gamma(x)=\sum_{N=1}^\infty
\left\langle Z_N(\beta|\chi) \right\rangle x^N$, is then given by
\begin{equation} \label{eq:GF:annealed}
\begin{split}
  \Gamma(x) = & \left[ \mathbf{f}^\mathrm{T} \mathbf{P}^{-1} \mathbf{\Gamma}_T (x)
  + (\mathbf{Wf})^\mathrm{T} \mathbf{R}^{-1} \right] \cdot \mathbf{\Gamma}_S (x) \\
  & \times \left[\mathbf{E} - \mathbf{\Gamma}_L (x) \mathbf{\Gamma}_S (x)\right]^{-1}
  \cdot \left[ \mathbf{\Gamma}_T (x) + \mathbf{E} \right] \cdot \mathbf{e}.
\end{split}
\end{equation}
Details of the calculation are given in the Appendix
\ref{app:GF:annealed}.  Here $\mathbf{f}=(f_A, f_B)^\mathrm{T}$ is
the vector of single monomer probabilities, $\mathbf{P}$ is the
transition probability matrix, Eq.~(\ref{eq:P:def}), the matrix
$\mathbf{R}$ is defined as $\mathbf{R}=\mathbf{PW}$ with the
``interaction matrix''
\begin{equation}
  \mathbf{W}=
  \left(\begin{array}{cc}
    w_A & 0 \\
    0  & w_B
  \end{array} \right) \, , \quad
  w_A\equiv e^{-\beta \varepsilon_A} \, , \quad
  w_B\equiv e^{-\beta \varepsilon_B} \, ,
\end{equation}
$\mathbf{E}$ is the unity matrix, and $\mathbf{e}$ is the vector
$\mathbf{e}=(1, 1)^\mathrm{T}$. The functions
$\mathbf{\Gamma}_S(x)$, $\mathbf{\Gamma}_L(x)$, and
$\mathbf{\Gamma}_T(x)$ are the GFs of adsorbed sequences, loops, and
tails, respectively, in matrix form,
\begin{equation} \label{eq:GFSLT:matrix}
\begin{split}
 & \mathbf{\Gamma}_S (x) = \sum_{n=1}^\infty \Omega_S(n) (\mathbf{PW})^{n} x^{n} = \sum_{n=1}^\infty \Omega_S(n) \mathbf{R}^{n} x^{n} \\
  & \mathbf{\Gamma}_L (x) = \sum_{n=1}^\infty \Omega_L(n) \mathbf{P}^{n} x^{n} \, , \, \mbox{ and }
  \mathbf{\Gamma}_T (x) = \sum_{n=1}^\infty \Omega_T(y) \mathbf{P}^{n} x^{n},
\end{split}
\end{equation}
where $\Omega_i(n)$ is the number of conformations corresponding to
each sequence of the type $i$ of length $n$. We note that
$\Gamma(x)$ is still a scalar function. The expression for
$\Gamma(x)$ has the same structure as that obtained by Birshtein for
homopolymer adsorption in Ref.~[\onlinecite{Birshtein:1979}],
except that it is a matrix equation in the RC case.

To find the smallest singularity of $\Gamma(x)$, one needs to
consider the smallest singularity associated with $\left[\mathbf{E}
- \mathbf{\Gamma}_L (x) \mathbf{\Gamma}_S (x)\right]^{-1}$ which can
be found as the smallest root of the equation
\begin{equation} \label{eq:det:annealed}
    \det\left[\mathbf{E} - \mathbf{\Gamma}_L (x) \mathbf{\Gamma}_S (x) \right]=0.
\end{equation}
It must then be compared with the smallest singularity $x_V$ of the
GF for the free chain in a bulk (in the absence of a surface),
\begin{equation}
 \label{eq:GFV:def}
 \Gamma_V(x) = \sum_{n=1}^{\infty} \Omega_V(n) \: x^n.
\end{equation}
Since the GFs of matrix arguments in Eq.~(\ref{eq:det:annealed}) are
matrix power series, they can be easily calculated using eigenvalues
and eigenvectors of the corresponding
matrices~\cite{Gantmacher:1959:1} ($\mathbf{P}$ and $\mathbf{R}$).
Despite its simplicity, Eq.~(\ref{eq:det:annealed}) can be solved
only numerically. The only case when it can be simplified
corresponds to Bernoullian, i.e. uncorrelated, copolymer, where
$c=0$. Applying the above scheme, it is straightforward to show that
the determinant equation (\ref{eq:det:annealed}) reduces to the
scalar one
\begin{equation} \label{eq:sing:annealed:uncorrel}
  1-\Gamma_L(x)\Gamma_S(x w_{eff}) = 0
\end{equation}
The effective $w_{eff} = f_A w_A + (1-f_B) w_B$, in accordance with
earlier finding~\cite{Soteros:2004}.

The method of GF allows the calculation of various properties of the
adsorbed chain. One of the most important quantitative
characteristics of polymer adsorption is the fraction of monomers in
the surface layer. Using the definition of $\left\langle
Z_n(\beta|\chi) \right\rangle$, Eq. (\ref{eq:Z:def}) and
differentiating $x_c$ with respect to $w_A$ and $w_B$, one can
calculate the fraction of A- and B-contacts with the surface:
\begin{equation} \label{eq:rho_AB:def}
  q_{A,B}=- \, \frac{\partial \log x_c}{\partial (\beta\varepsilon_{A, B})} = -\frac{w_{A, B}}{x_c} \cdot
  \frac{\partial x_c}{\partial w_{A, B}}
\end{equation}
The derivatives of $x_c$ with respect to $w_{A,B}$ are obtained from
the derivatives of lhs of Eq.~(\ref{eq:det:annealed})
\begin{equation} \label{eq:xc:def}
  \frac{\partial x_c}{\partial w_{A, B}} = -\frac{\partial D/\partial w_{A, B}}{\partial D/\partial x}\Biggr|_{x_c} ,
\end{equation}
where we have defined
\begin{equation}
  D:= \det\left[\mathbf{E} - \mathbf{\Gamma}_L (x) \mathbf{\Gamma}_S (x) \right] .
\end{equation}
To differentiate the determinant $D$ we use Jacobi's formula
\begin{equation}
  \frac{d \det \mathbf{X}(\tau)}{d\tau} = \Trace \left(\Adj (\mathbf{X})
   \cdot \frac{d\mathbf{X}}{d\tau} \right)
\end{equation}
where ``Trace'' stands for the trace of a matrix and $\Adj
(\mathbf{X})$ denotes the adjugate matrix for $\mathbf{X}$. Hence,
\begin{equation}
      \frac{\partial D}{\partial x} =
      - \Trace \left[\Adj (\mathbf{E} - \mathbf{\Gamma}_L (x) \mathbf{\Gamma}_S (x)) \cdot
      \left( \frac{\partial \mathbf{\Gamma}_L}{\partial x} \cdot \mathbf{\Gamma}_S (x) +
      \mathbf{\Gamma}_L (x) \cdot \frac{\partial \mathbf{\Gamma}_S}{\partial x} \right) \right],
\end{equation}
and
\begin{equation}
  \frac{\partial D}{\partial w_{A,B}} = - \Trace \left[\Adj (\mathbf{E} - \mathbf{\Gamma}_L (x) \mathbf{\Gamma}_S (x)) \cdot
   \mathbf{\Gamma}_L (x) \cdot \frac{\partial \mathbf{\Gamma}_S}{\partial w_{A,B}} \right].
\end{equation}

After having derived the appropriate formalism for annealed RC
adsorption, we are now ready to generalize our results to the Morita
approximation. The latter differs from the annealed approximation in
that it imposes conditions on the first and second moments of
monomers' distribution, Eq.~(\ref{eq:MConstr:deriv}).
Comparing the expressions for the
partition function, Eqs~(\ref{eq:Za:def}) and (\ref{eq:ZM2:def}), it
is easy to see that the calculation of the GF in the Morita
approximation $\Gamma(x)=\sum_N \left\langle Z_N(\beta|\chi)
\right\rangle_M x^N$ is equivalent to that in the annealed
approximation, if the following changes are made: $x\to x
e^{-\lambda f_A - \kappa f_{AA}}$ and
\begin{equation} \label{eq:P:Morita}
   \mathbf{P}=
   \left(
      \begin{array}{cc}
         p_{AA} e^{\lambda + \kappa} & p_{AB} \\
         p_{BA} e^{\lambda} & p_{BB}
      \end{array}
   \right).
\end{equation}
The matrix $\mathbf{P}$ is again $\mathbf{R}=\mathbf{PW}$. If we
denote $y:=x e^{-\lambda f_A - \kappa f_{AA}}$, we obtain the analog
of Eq.~(\ref{eq:det:annealed})
\begin{equation} \label{eq:det:Morita}
    \det\left[\mathbf{E} - \mathbf{\Gamma}_L (y) \mathbf{\Gamma}_S (y) \right]=0.
\end{equation}

To find the values of the variational parameters $\lambda$ and
$\kappa$, the Morita conditions (\ref{eq:MConstr:deriv}) are
applied. In terms of $y$ they can be rewritten as
\begin{equation} \label{eq:MConstr:y}
  \begin{split}
      & \frac{\partial \log y}{\partial \lambda} + f_A = 0 \, , \\
      & \frac{\partial \log y}{\partial \kappa} + f_{AA} = 0
  \end{split}
\end{equation}
where the logarithmic derivatives of $y$ can be expressed by
differentiating Eq.~(\ref{eq:det:Morita})  with respect to $\lambda$
and $\kappa$. By solving equations (\ref{eq:MConstr:y}) together
with Eq.~(\ref{eq:det:Morita}), the values of $y_c$, $\lambda$ and
$\kappa$ and, therefore, the smallest singularity $x_c=y_c
e^{\lambda f_A + \kappa f_{AA}}$, as functions of $\beta$,
$\varepsilon_A$ and $\varepsilon_B$ are obtained.

The calculation of observables is similar to that described for the
annealed case with the only change $x_c \rightarrow y_c$. Indeed,
$x_c = x_c[w_A, w_B, \lambda(w_A, w_B), \kappa(w_A, w_B)]$ and
\begin{equation}
    \frac{d \log x_c}{d w_A} = \frac{\partial \log x_c}{\partial w_A}
       + \frac{\partial \log x_c}{\partial \lambda}\cdot \frac{\partial \lambda}{\partial w_A}
       + \frac{\partial \log x_c}{\partial \kappa}\cdot \frac{\partial \kappa}{\partial w_A}
\end{equation}
Since ${\displaystyle \frac{\partial \log x_c}{\partial \lambda} }$
and ${\displaystyle \frac{\partial \log x_c}{\partial \kappa} }$ are
equal to zero because of the Morita conditions, we get
\begin{equation}
    \frac{d \log x_c}{d w_A} = \frac{\partial \log x_c}{\partial w_A} =
  \frac{\partial \log y_c}{\partial w_A}
\end{equation}
and the same is, of course, valid for the ${\displaystyle \frac{d
\log x_c}{d w_B} }$.

The final missing ingredient in the theory is the actual form of the
GFs of loops and adsorbed sequences, $\Gamma_S(x)$ and
$\Gamma_L(x)$. They depend on the details of the particular system
under consideration, {\em i.e.}, on the lattice type and on the
geometry of the adsorbing substrate. Here we consider the case of
the adsorption on a plane, when polymer conformations are simple
random walks, restricted to the half-space, on the conventional
6-choice simple cubic lattice (SCL).

Calculating the GF for adsorbed sequences is a relatively
simple task. Using the definition of the GF (\ref{eq:GF:def}), we obtain:
\begin{equation} \label{eq:GS:plane:6scl}
  \Gamma_S(x)=x + (z-2)\,x^2 + (z-2)^2\,x^3 + \ldots = \frac{x}{1-(z-2)x}
\end{equation}
Here $z=6$ is the coordination number of the SCL.

The calculation of the GF for loops is more complicated.
It is described in Appendix \ref{app:GL}. The result is
\begin{equation}
  \Gamma_L(x)=\frac{1}{2x}\left\{ 1- (z-2)x - \sqrt{(1-zx)[1-(z-4)x]} \right\} \, .
\end{equation}

%
\subsection{Numerical simulations}
\label{sec:numerics}

Our analytical calculations are complemented by numerical
simulations of RC adsorption on planar surfaces. The numerical
method is described below. The model is the same (section
\ref{sec:model_def}, but the system under consideration is slightly
different. First, the analytical calculations were carried out for
chains of infinite length, whereas in the numerical simulations, the
chains must remain finite for obvious reasons. The main consequence is that
we no longer have a sharp adsorption transition, but a smooth
crossover region between an adsorbed and a desorbed regime. We have
not studied chain-length effects in this work. Second, the chains in
the numerical simulations are tethered to the surface at one end.
This helps to avoid an uncertainty with the normalization, because
for free chains of finite length one would have to introduce a box
of finite size. We note that tethered chains and free chains have
the same adsorption characteristics in the infinite-chain limit,
hence our analytical results also apply to tethered chains (in our
theory, the main equation (\ref{eq:det:annealed}) or Eq.
(\ref{eq:det:Morita}) only contains GFs for adsorbed sequences and
loops and is therefore equally valid for both free and tethered
chains).

The numerical treatment is based on a Green's functions formalism
first introduced by Rubin \cite{Rubin:NBS:1965, Rubin:NBS:1966,
Rubin:JCP:1965} and later used in the more general theory of
Scheutjens and Fleer \cite{Fleer:1993}. The central quantities are
the statistical weights $G_t(z;n)$ of all conformations of tethered
RC chain parts of length $n$ with one free end in the layer $z$, and
the corresponding weights $G_f(z;N-n)$ of all conformations of chain
parts between the $n$th monomer and the end monomer $N$, subject to
the constraint that the monomer $n$ is fixed in the layer $z$
whereas the end monomer is free. They satisfy the recurrent
relations
\begin{equation}
\label{recursion} G_{t,f}(z;n+1) = \left\{ \lambda G_{t,f}(z-1;n) +
(1-2\lambda)G_{t,f}(z;n) + \lambda G_{t,f}(z+1;n) \right\}
\end{equation}
at $z \ne 0$ and
\begin{equation}
\label{recursion0} G_t(0;n+1) = \exp(-\beta \varepsilon_{n+1})
\left\{ (1-2\lambda)G_t(0;n) + \lambda G_t(1;n) \right\},
\end{equation}
\begin{equation}
\label{recursion1} G_f(0;n+1) = \exp(-\beta \varepsilon_{N-n+1})
\left\{ (1-2\lambda)G_t(0;n) + \lambda G_t(1;n) \right\}
\end{equation}
with $\varepsilon_n = \chi_n \varepsilon_A +
(1-\chi_n)\varepsilon_B$,  where $\lambda$ is the probability that a
random walk step connects neighboring layers. On the simple cubic
lattice, one has $\lambda=\frac{1}{6}$. Using as the starting point
the monomer segment distribution $G_t(0;1) = \exp(-\beta
\varepsilon_1)$, $G_f(0;1) = \exp(- \beta \varepsilon_N)$, and
$G_t(z;1)=0$, $G_f(z;1)=1$ for $z > 0$, and recursively applying
Eqs.~(\ref{recursion}--\ref{recursion1}), one can easily calculate
$G_{t,f}(z;n)$ for every chain length $n$. The statistical weight of all
conformations of tethered chain is then obtained by summing over
all positions of the free end:
\begin{equation}
\label{Z_teth} Z(N)=\sum_{z=0}^{Nl} G_t(z;N)
\end{equation}

The change in the free energy of the tethered chain with respect to
the free chain in the solution is given by $ \Delta F=-\log Z(N) $.
Here the translational entropy of the free chain has been
disregarded, {\em i.e.}, the chains are assumed to be sufficiently
long that it can be neglected. At the transition point one has
$\Delta F=0$, {\em i.e.}, the energetic benefit of monomer-surface
contacts is equal to the entropic penalty caused by tethering the
chain to the plane and thereby restricting the number of possible
conformations.

Using $G_t$ and $G_f$ one can also calculate the probability that
monomer $n$ is adsorbed on the surface via the composition law
\begin{equation}
    p(n) = \frac{G_t(0,n) G_f(0,N-n) e^{\beta \varepsilon_n}}{Z(N)}.
\end{equation}
This gives the average fraction of adsorbed monomers
\begin{equation}
    q = \frac{1}{N}\sum_{n=1}^N p(n),
\end{equation}
and the average fractions of A and B contacts with the surface
\begin{equation}
    q_A = \frac{1}{N}\sum_{n=1}^N p(n) \delta_{\chi_i, 1} \, , \quad
    q_B = \frac{1}{N}\sum_{n=1}^N p(n) \delta_{\chi_i, 0} = q - q_A.
\end{equation}

In this paper, we present results for chains of length $N=1000$. We
consider two-letter correlated RC sequences with the distribution
defined in Section \ref{sec:model_def}. For every set of model
parameters $f_A$ and $c$, the adsorption characteristics were
calculated as a function of the inverse temperature $\beta$ for 100
different sequence realizations and then averaged. This corresponds
to a situation with \emph{quenched} sequence disorder.

\section{Results} \label{sec:results}
\subsection{Annealed chains} \label{subs:annealed}

We first discuss the adsorption properties of annealed chains,
focussing on the results for the two cases: sticky-neutral (SN) and
sticky-repulsive (SR) defined in Sec.~\ref{sec:model_def}.

\begin{figure}[t]
  \begin{center}
  \includegraphics[width=7.4cm]{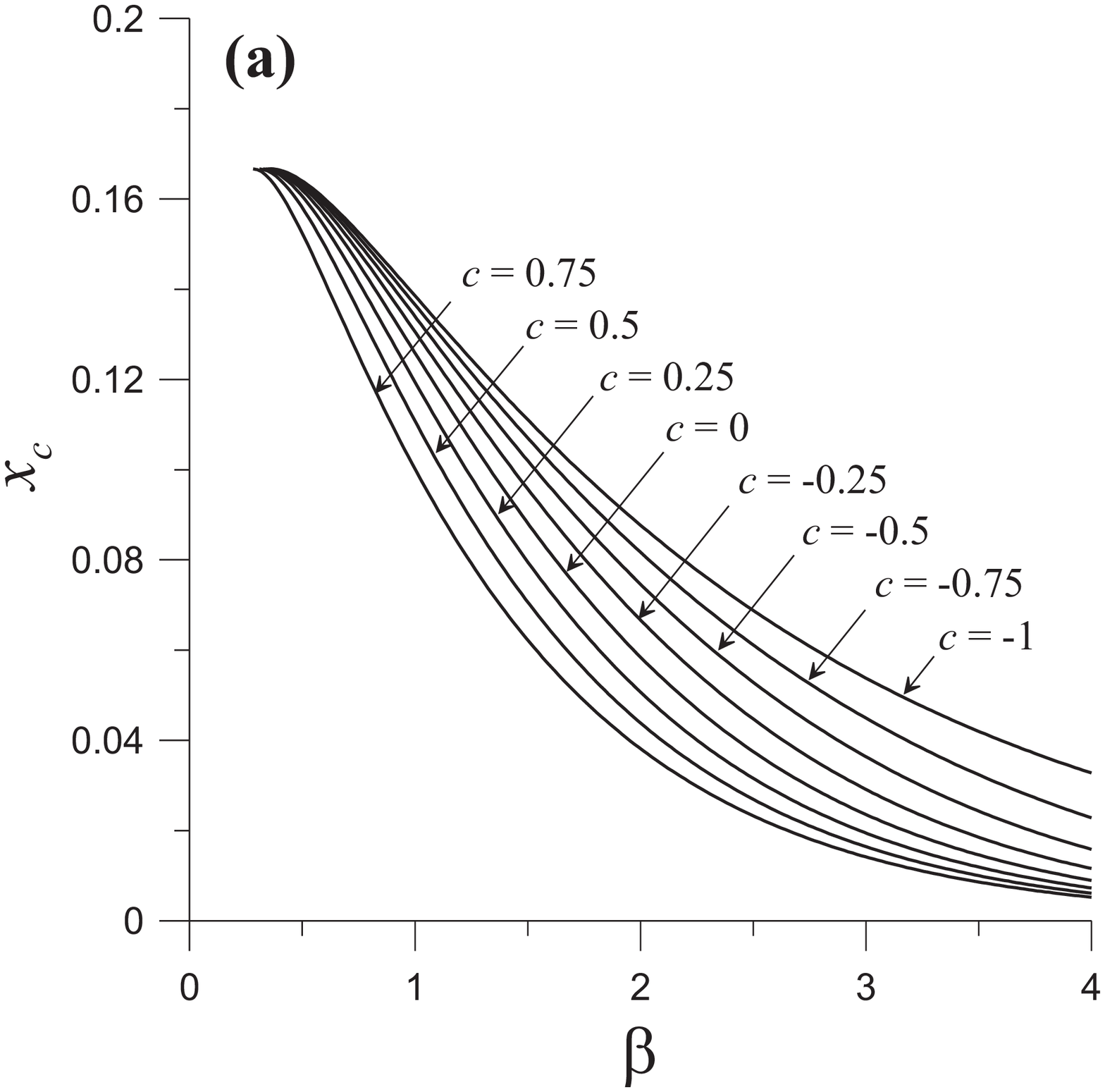}
  \includegraphics[width=7.4cm]{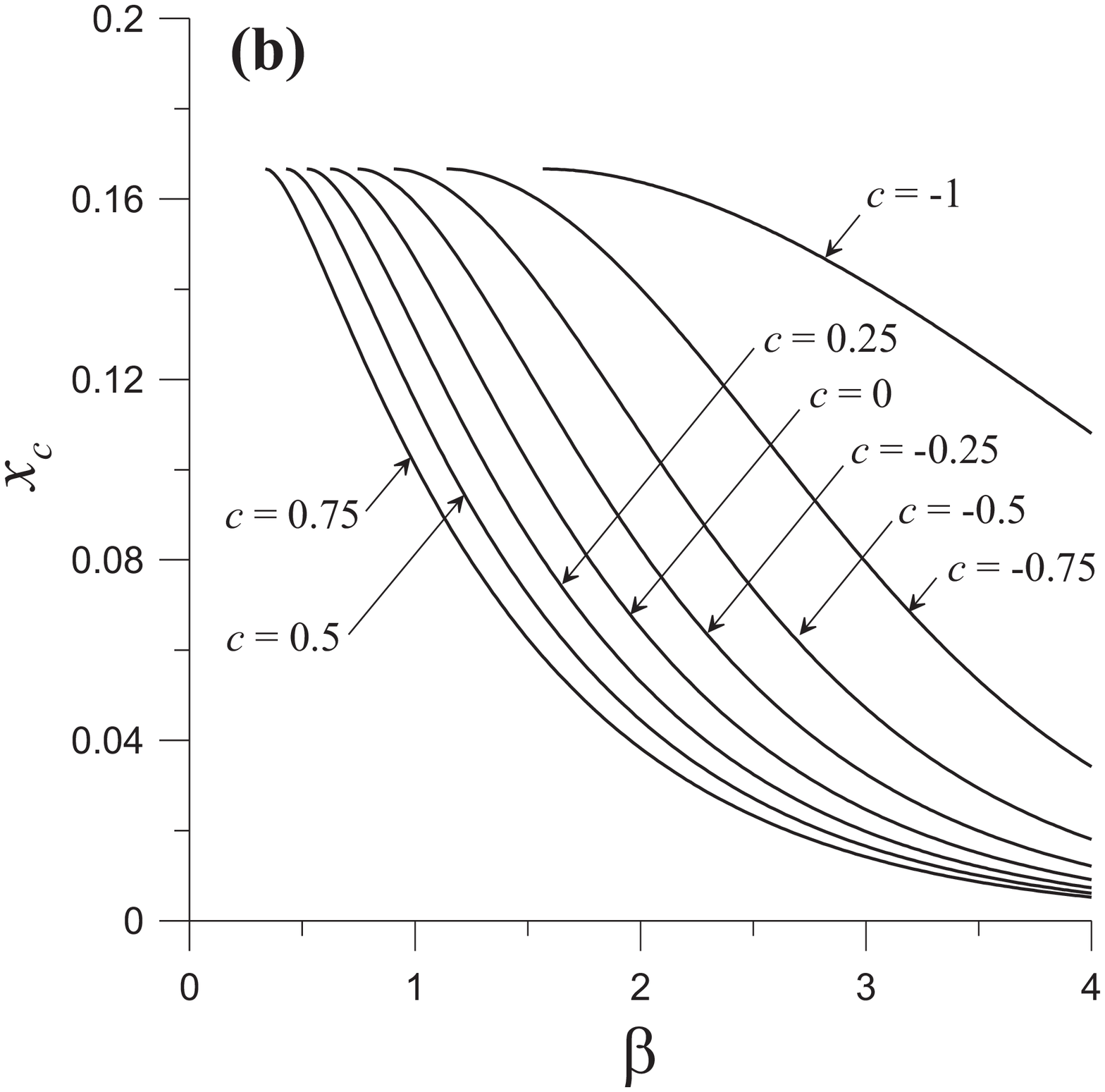}
  \end{center}
  \caption{Solution of Eq.~(\ref{eq:det:annealed}) for $f_A=0.5$,  $\varepsilon_A=-1$, $\varepsilon_B=0$ (a) $1$ (b) 
  and different values of cluster parameters $c=-1$, $-0.75$, $-0.5$, $-0.25$, $0$, $0.25$, $0.5$, $0.75$.}
  \label{fig:ann:xc}
\end{figure}

\begin{figure}[t]
  \begin{center}
  \includegraphics[width=7.4cm]{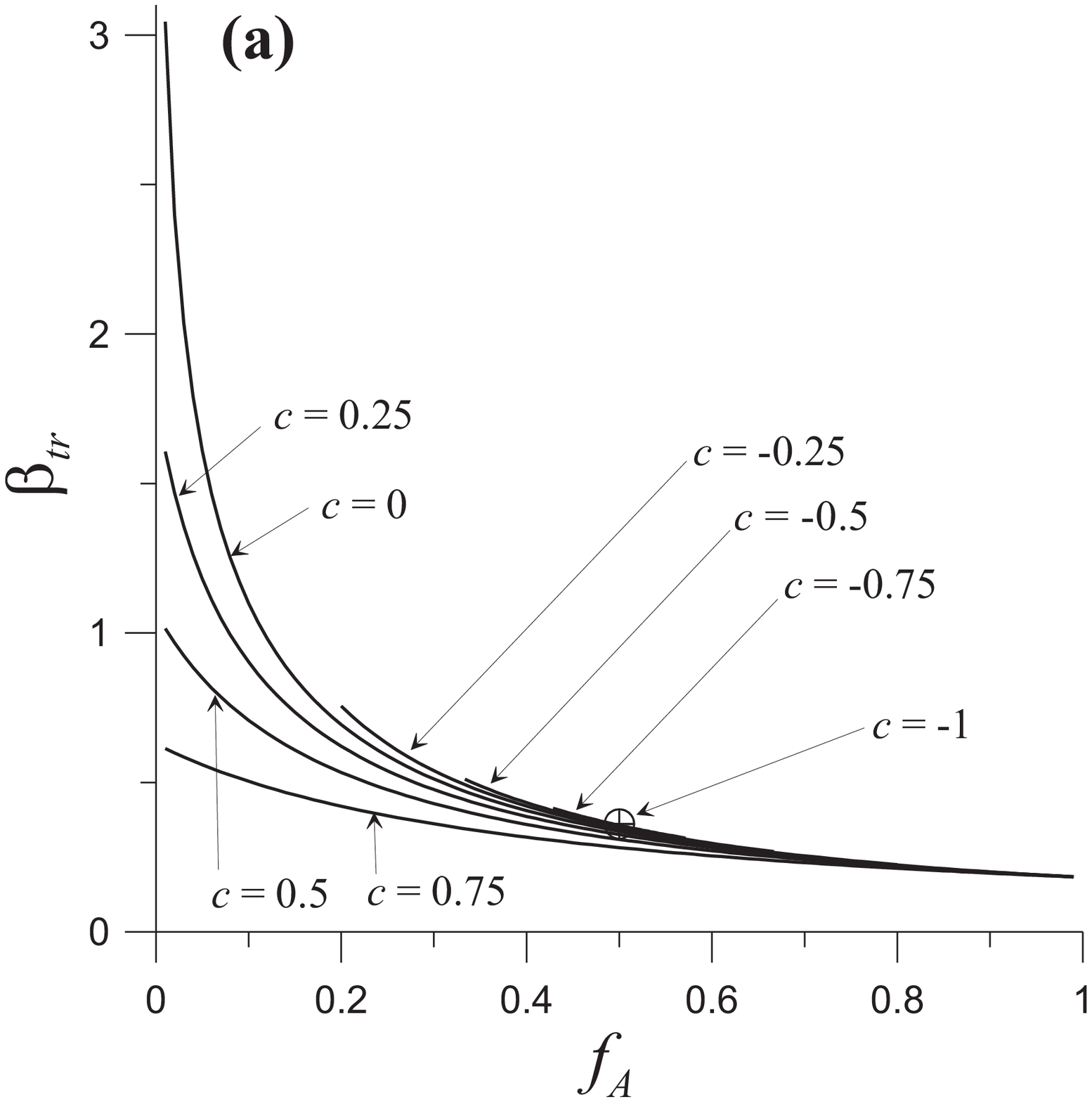}
  \includegraphics[width=7.4cm]{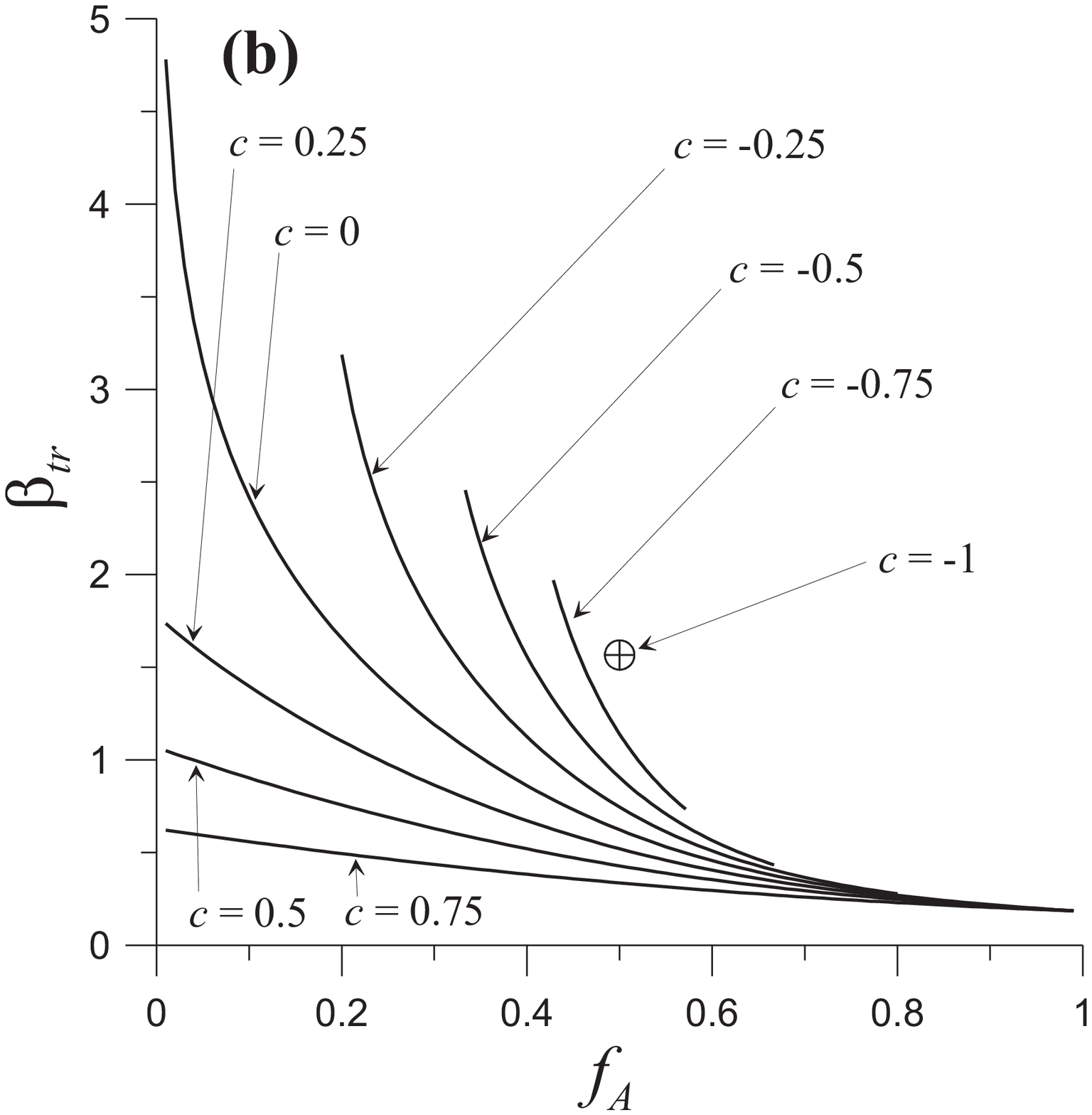}
  \end{center}
  \caption{Inverse adsorption transition temperature in the annealed approximation as function of A-monomer probability 
  for $\varepsilon_A=-1$, $\varepsilon_B=0$ (a) $1$ (b) and different values of cluster parameters 
  $c=-0.75$, $-0.5$, $-0.25$, $0$, $0.25$, $0.5$, $0.75$. The transition point for $f_A=0.5$ and $c=-1$ is shown 
  by symbol $\oplus$.}
  \label{fig:ann:tp}
\end{figure}

Fig.~\ref{fig:ann:xc} shows the results of the numerical solution of
Eq. (\ref{eq:det:annealed}) as a function of the inverse temperature
$\beta$ at $f_A = 0.5$ for various values of $c$ in the SN and SR
cases. One can see that blockier RCs with higher values of $c$ have
lower $x_c$ (hence, a lower free energy $\beta F/N =\log x_c$). At
small values of $\beta$ (at high temperatures), all the curves
terminate in the adsorption-desorption transition point where $x_c =
x_V$ (recall that $x_V$ is the smallest singularity of $\Gamma_V(x)$
-- the GF for the bulk chain, Eq.~(\ref{eq:GFV:def}); for a simple
cubic lattice one has $x_V = 1/6$). The transition point, the
critical value $\beta_{tr}$ of $\beta$, is therefore found from the
equation
\begin{equation} \label{eq:tp:annealed}
    \det\left[\mathbf{E} - \mathbf{\Gamma}_L (x_V) \mathbf{\Gamma}_S (x_V) \right]=0.
\end{equation}
The resulting value of $\beta_{tr}$ as a function of $f_A$, the
relative weight of $A$-monomers in the sequence, is shown in
Fig.~\ref{fig:ann:tp}. For $c<0$ the allowed range of $f_A$ 
narrows according to Eq.~(\ref{eq:fa:range}). In the case $c=-1$ 
it degenerates to a single point $f_A=0.5$. It can be seen
that the aggregation of monomers into clusters shifts the transition
points and the $x_c(\beta)$ curve towards lower values of $\beta$
({\em i.e.}, it favors adsorption) and this shift is more pronounced
in the SR case.

The equation for the transition point (\ref{eq:tp:annealed}) has
some similarity with the result obtained by Brun~\cite{Brun:1999}.
However, the latter contains only the largest eigenvalues,
$\nu_{max}$, $\mu_{max}$, of the matrices $\mathbf{P}$ and
$\mathbf{R}$, respectively. In our notation the equation of Brun has
the form: $1-\Gamma_L(\nu_{max} x_V)\Gamma_S(\mu_{max} x_V) = 0$. It
gives the same result as Eq.~(\ref{eq:tp:annealed}) in the
Bernoullian case ($c=0$), but overestimates $\beta_{tr}$ at $c<0$
and underestimates it at $c>0$, compared to our result.

\begin{figure}[t]
  \begin{center}
  \includegraphics[width=7.4cm]{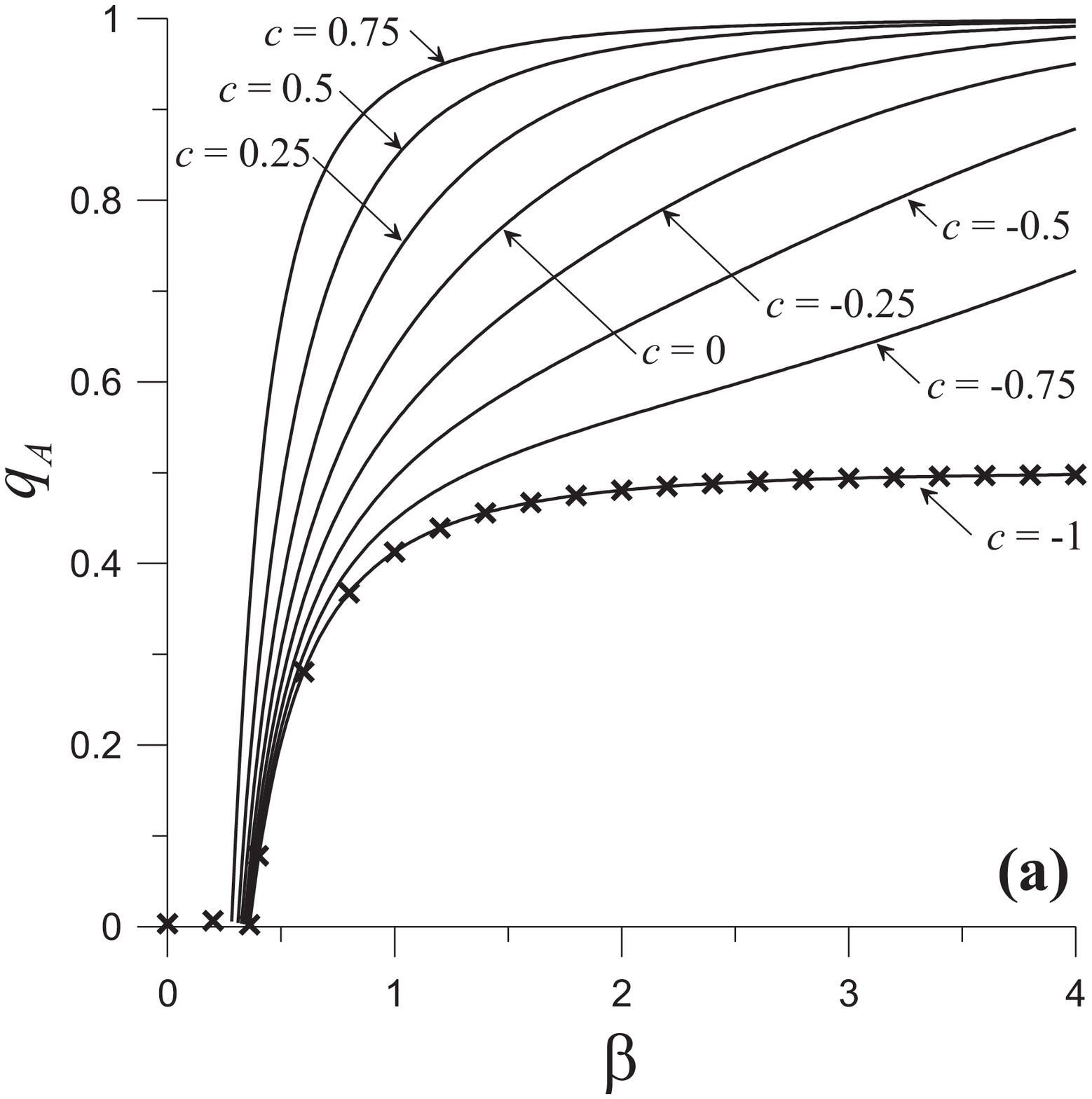}
  \includegraphics[width=7.4cm]{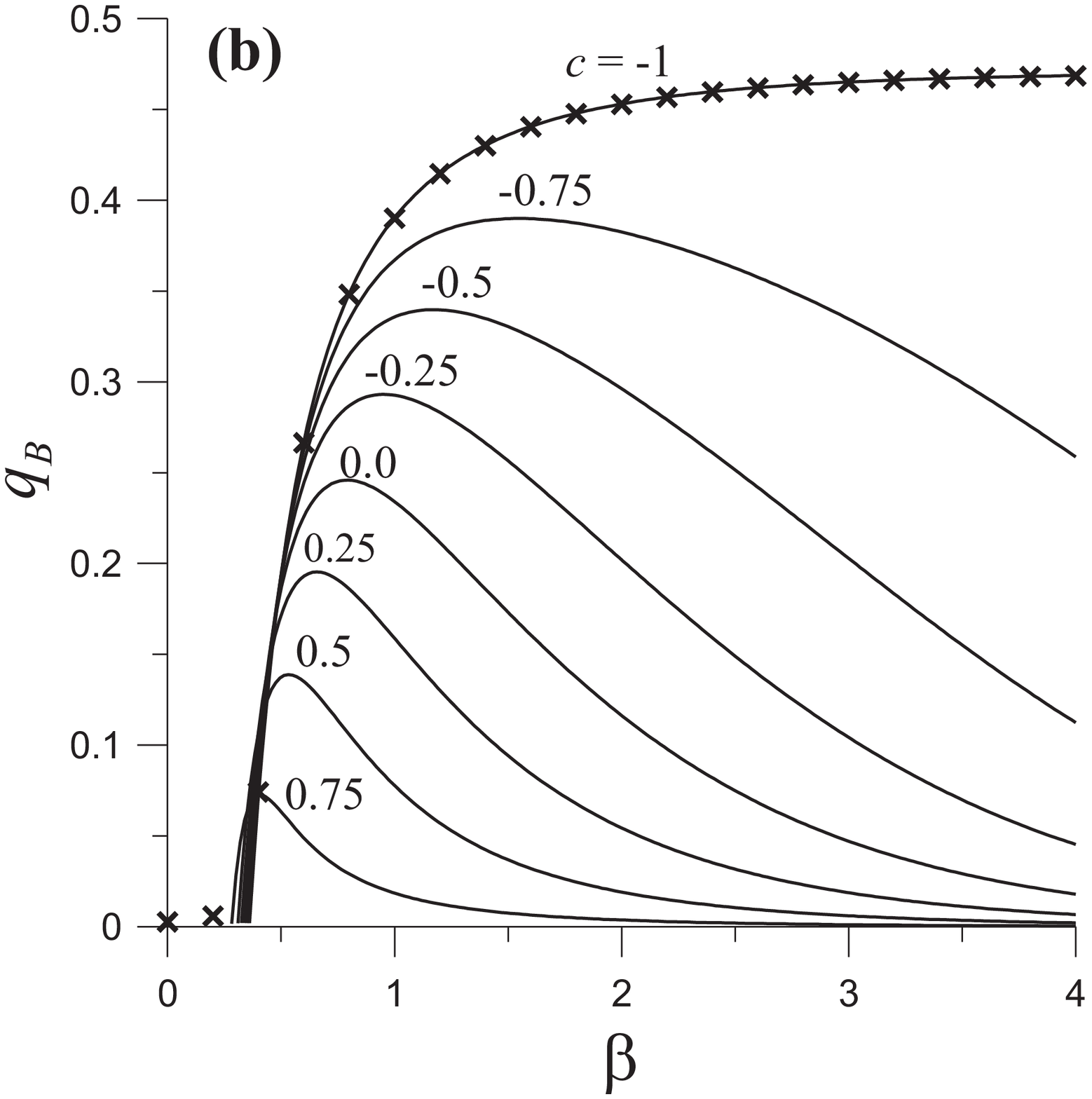}
  \end{center}
  \caption{Fraction of A and B contacts in the annealed approximation for $f_A=0.5$,  $\varepsilon_A=-1$, 
  $\varepsilon_B=0$ and different values of cluster parameters 
  $c=-1$, $-0.75$, $-0.5$, $-0.25$, $0$, $0.25$, $0.5$, $0.75$ as indicated (lines). 
  Also shown are numerical results for $c=-1$ (crosses).}
  \label{fig:ann:sn}
\end{figure}

\begin{figure}[t]
  \begin{center}
  \includegraphics[width=7.4cm]{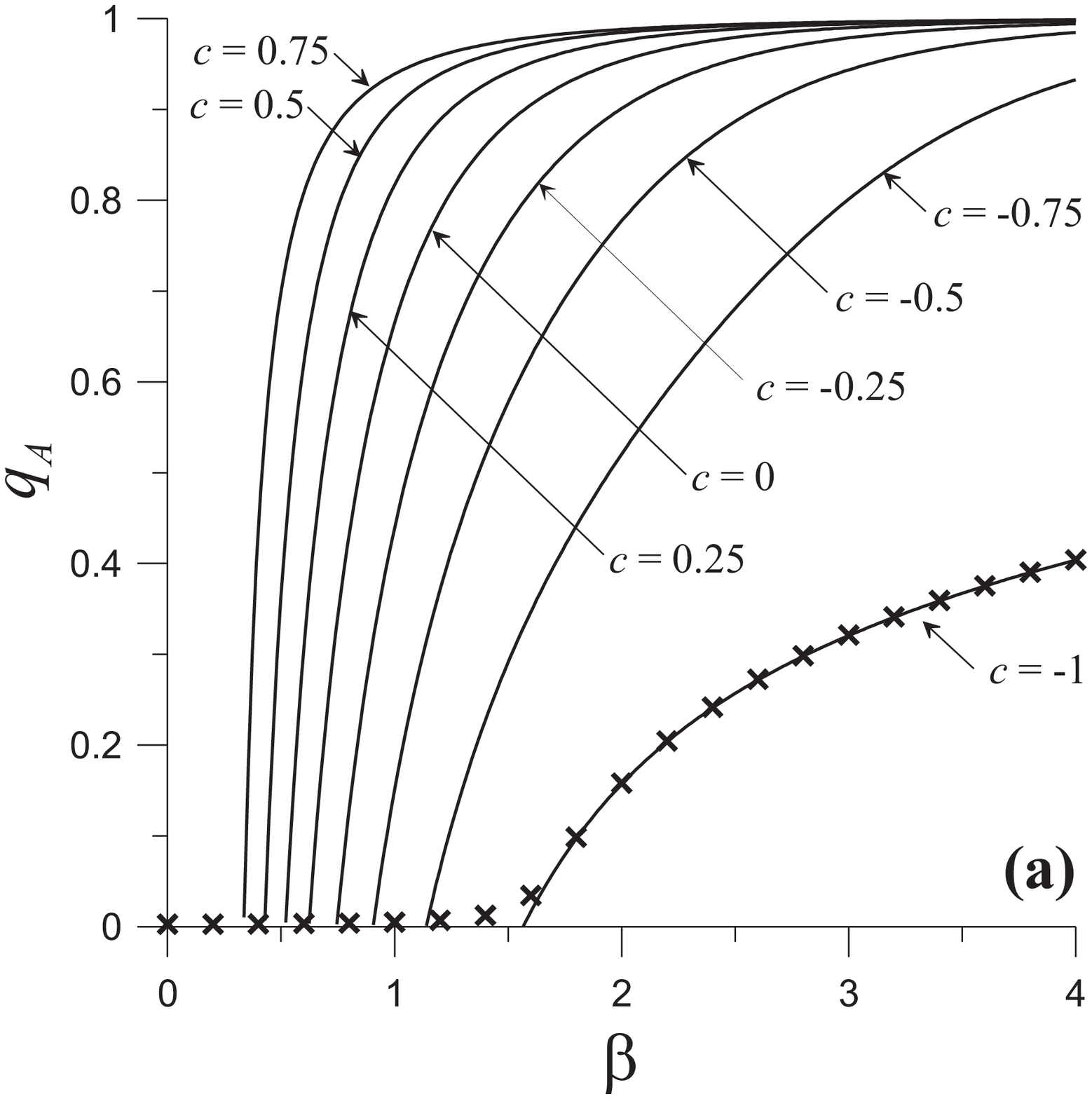}
  \includegraphics[width=7.4cm]{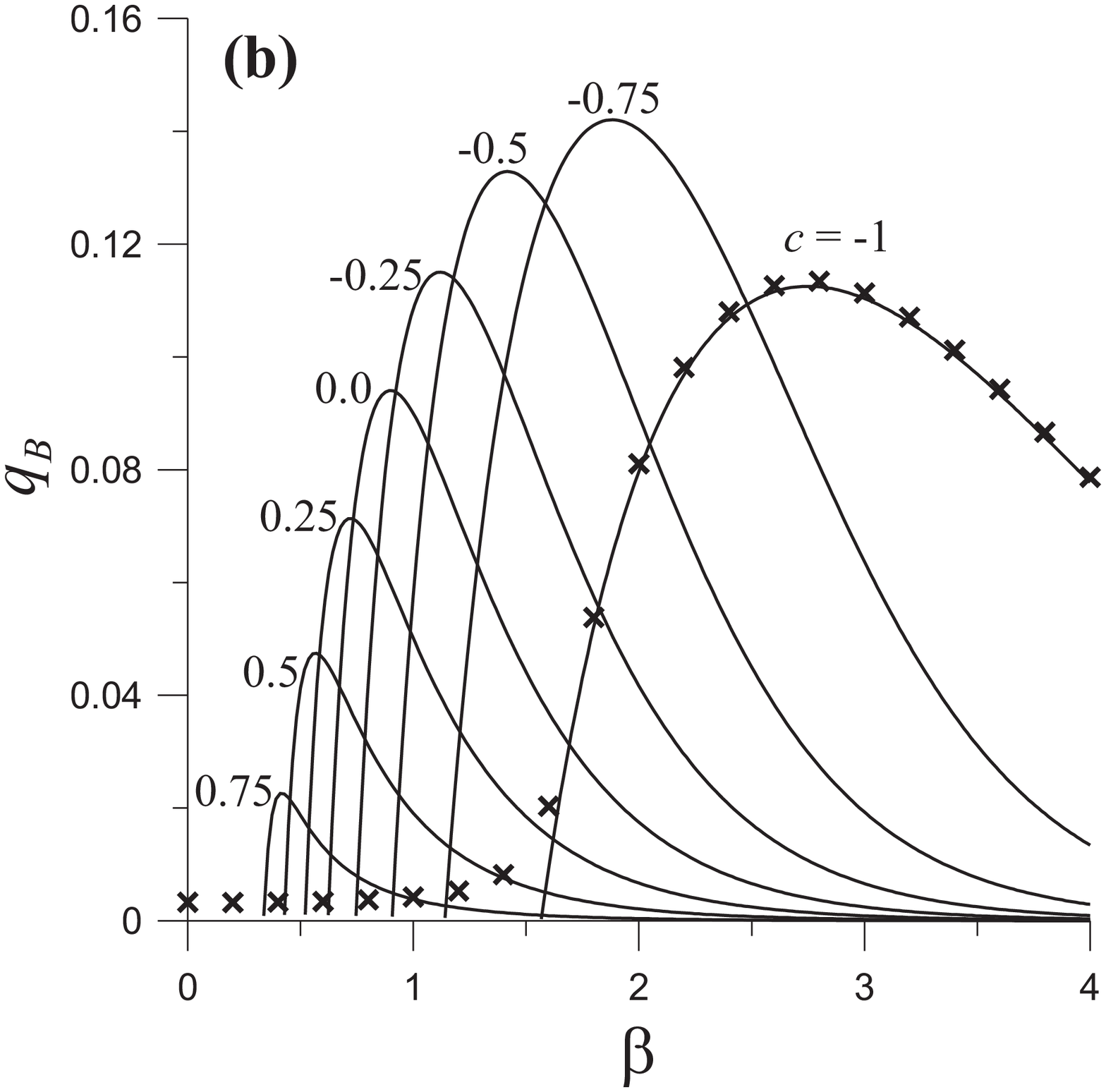}
  \end{center}
  \caption{Fraction of A and B contacts in the annealed approximation for $f_A=0.5$,  $\varepsilon_A=-1$, 
  $\varepsilon_B=1$ and different values of cluster parameters 
  $c=-1$, $-0.75$, $-0.5$, $-0.25$, $0$, $0.25$, $0.5$, $0.75$ as indicated (lines). 
  Also shown are numerical results for $c=-1$ (crosses).}
  \label{fig:ann:sr}
\end{figure}

Figs~\ref{fig:ann:sn} and \ref{fig:ann:sr} show examples of the
temperature dependence of the fraction of A- and B-contacts for the
two cases SN and SR. 

\begin{figure}[t]
  \begin{center}
  \includegraphics[width=7.4cm]{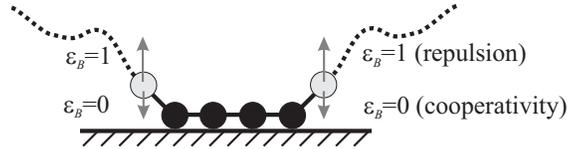}
  \end{center}
  \caption{Illustration of cooperativity effect: Adsorbed train and its adjacent monomers.}
  \label{fig:cooperativity}
\end{figure}

At $c=-1$, where the chain is a regular alternating copolymer, 
the ''annealed approximation'' of course becomes exact and the 
results are in excellent agreement with the corresponding
numerical results, which are also shown for comparison. The remaining 
difference is due to the finite chain length in the numerical
calculations. This situation can be used to illustrate the main 
qualitative difference between the two cases SN and SR:
The fraction of B-contacts (Figs. \ref{fig:ann:sn} b and
\ref{fig:ann:sr} b) behaves differently depending on whether
B-monomers are neutral or repelling. In the former case $q_B$ grows
with $\beta$ whereas in the latter case it behaves nonmonotonically.
The initial growth is caused by a cooperativity effect: 
B-monomers linked to adsorbed A-monomers have a higher
probability to come into contact with the surface (see 
Fig.~\ref{fig:cooperativity}). In the 
SN case, such contacts are not penalized, hence the B-monomers
are effectively attracted by the surface and $q_B$ continues
to grow for all $\beta$. In the SR case, the cooperativity
effect is counterbalanced by the increasing effect of the energy
penalty on B-contacts, hence $q_B$ decreases again at high $\beta$.

Next we consider the curves corresponding to truly disordered chains 
with $c > -1$. Both in the SN and SR cases the fraction of A-contacts
grows with increasing $\beta$ up to $q_A=1$, which is much higher
than the expected fraction $f_A$ of A-stickers \emph{in the
sequence} (not necessarily adsorbed) for quenched disorder ($q_A
(\beta \gg 1) \le f_A$=0.5). At the same time, the fraction of
B-contacts exhibits a nonmonotonic behaviour both in the SN and 
the SR case: Close to the transition point it grows, and then, after 
passing through a maximum, it decreases and virtually vanishes. 
This reflects the physical nature of the annealed approximation 
mentioned above: Any monomer can change its affinity to the surface 
(from negative to positive), if it is thermodynamically preferable 
for the system, and this is exactly the case at high $\beta$. 
Close to the transition point, the degree of A$\to$B conversion is 
small and neutral or repelling B-monomers come into contact with 
the surface due to the cooperativity effect described above; of 
course, for given chain statistic (given $f_A$ and $c$)
and given inverse temperature $\beta$, $q_B$ is smaller in the SR
case than in the SN case due to the additional penalty on
B-contacts.

This example demonstrates that the annealed approximation gives us
essentially incorrect result regarding one of the most important
conformational characteristics of adsorbed heteropolymer chains.
Therefore it is necessary to improve on this approach. The Morita
approximation (Sec.~\ref{sec:morita}) provides us with such a
possibility.

%

\subsection{'Morita chains' and Quenched chains} \label{subs:morita}

The objectives of the present work are not only to apply the Morita
approximation for the problem of adsorption of correlated RCs, but
also to estimate how good it can approximate the quenched (i.e. the
real) system. The results obtained using the Morita approximation
and the numerical approach (Sec.~\ref{sec:numerics}) are thus
presented together. We will use the following convention: The
results obtained with the Morita approximation are shown as solid 
lines, whereas the data for the corresponding quenched systems
are represented by symbols. Since our primary interest is 
to study the influence of correlations and the nature of the 
non-adsorbing monomer B on the RC adsorption
properties, we set $f_A = 0.5$ in this section. 
The results for $c=-1$ are not shown because they coincide with those 
obtained using the annealed approximation and have already been
presented in Figs~\ref{fig:ann:sn} and \ref{fig:ann:sr}.

\subsection{SN case} \label{sec:results:sn}
\begin{figure}[t]
  \begin{center}
  \includegraphics[width=7.4cm]{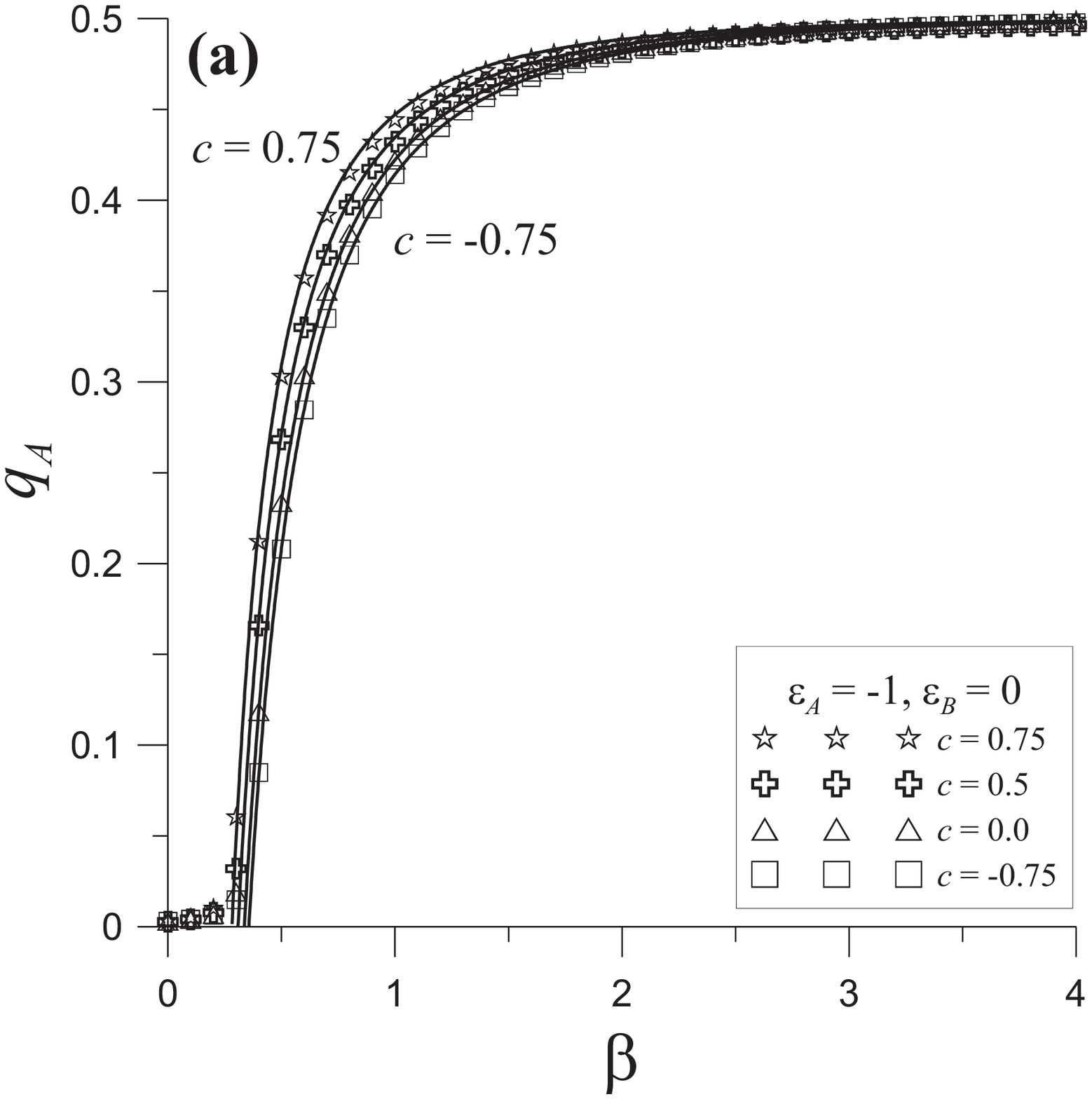}
  \includegraphics[width=7.4cm]{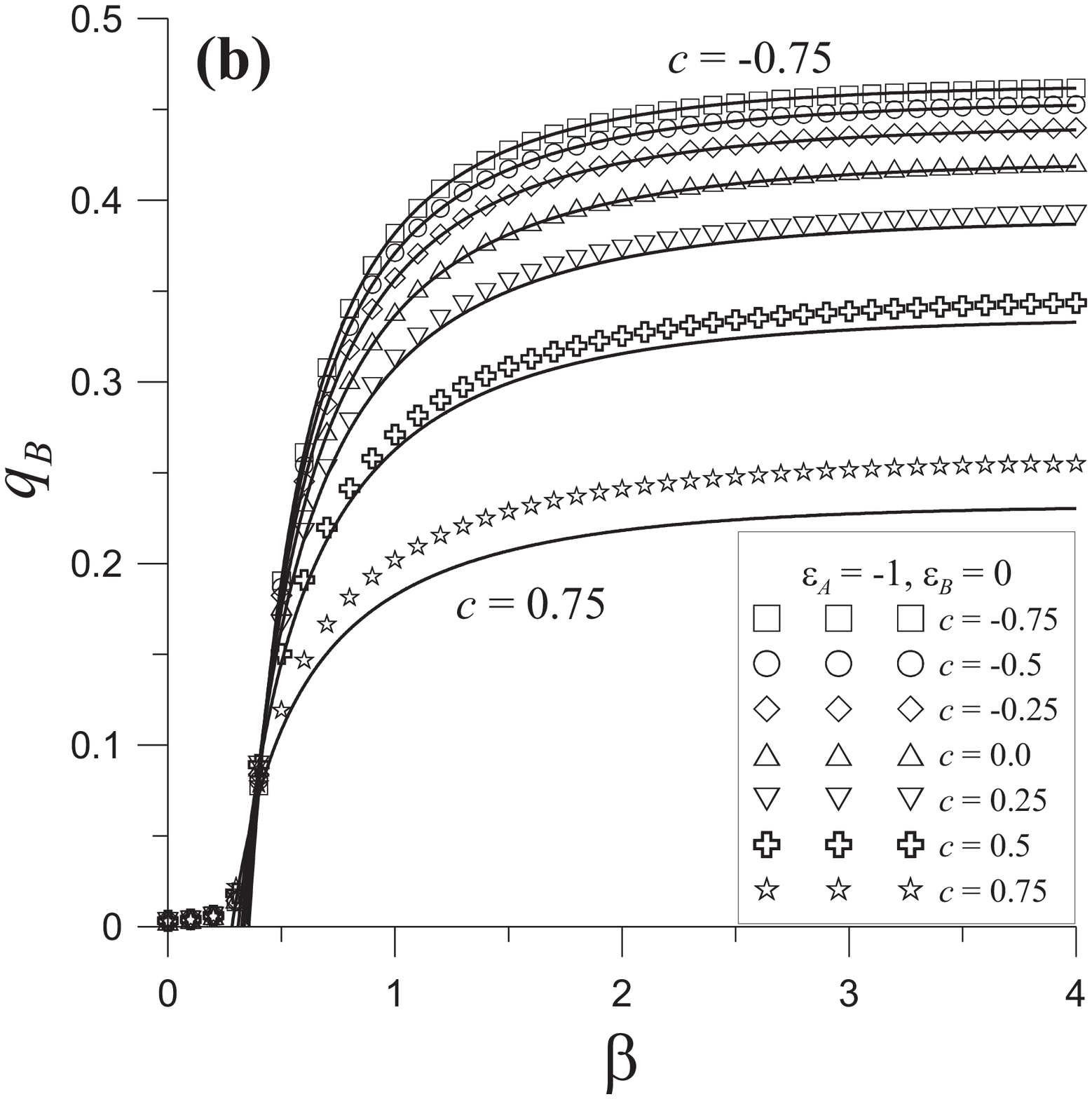}
  \end{center}
  \caption{Fraction of A and B contacts calculated in the Morita approximation (lines) and using the transfer matrix approach
  (symbols) for $f_A=0.5$,  $\varepsilon_A=-1$, $\varepsilon_B=0$ and different values of cluster parameters
  $c=-0.75$~($\Box$), $-0.5$~($\bigcirc$), $-0.25$~($\Diamond$), $0$~($\bigtriangleup$), $0.25$~($\bigtriangledown$),
  $0.5$~($+$), $0.75$~($\star$).}
  \label{fig:Mor:sn}
\end{figure}

We consider first the SN case. Fig.~\ref{fig:Mor:sn} shows the
temperature dependence of the fraction of A and B contacts for
different values of the cluster parameter $c$. Both $q_A$ and $q_B$
grow with increasing $\beta$. The results for $q_B$ differ
qualitatively from those obtained for all random copolymers
($c < -1$) in the annealed approximation, but they resemble those 
obtained for alternating copolymers ($c=-1$) in 
Fig.~\ref{fig:ann:sn}. The reason for the monotonic growth of 
$q_B$ is the absence of an energetic penalty on a B contact,
combined with the cooperativity effect discussed earlier
(Fig.~\ref{fig:cooperativity}). This effect explains the
larger amount of B-contacts for RCs with lower $c$ at intermediate
and strong adsorption (far from the transition point). Contrary to
the annealed case (Fig.~\ref{fig:ann:sn}~a), the value of $q_A$ does
not exceed $f_A$.

Comparing the results obtained using the Morita approximation and
the numerical procedure, we have not only qualitative, but also a
quantitative agreement. The curves for $q_A$ are predicted almost
perfectly by the Morita approximation, the results are only very
slightly overestimated. The $q_B$ curves are in reasonable agreement
for $c\leq 0$, whereas for $c>0$, the Morita approximation
underestimates $q_B$. This can be explained by the fact that the
Morita copolymer has the possibility to adjust itself to the surface
-- A and B monomers can be redistributed between trains and loops.

\subsection{SR case} \label{sec:results:sr}
\begin{figure}[t]
  \begin{center}
  \includegraphics[width=7.4cm]{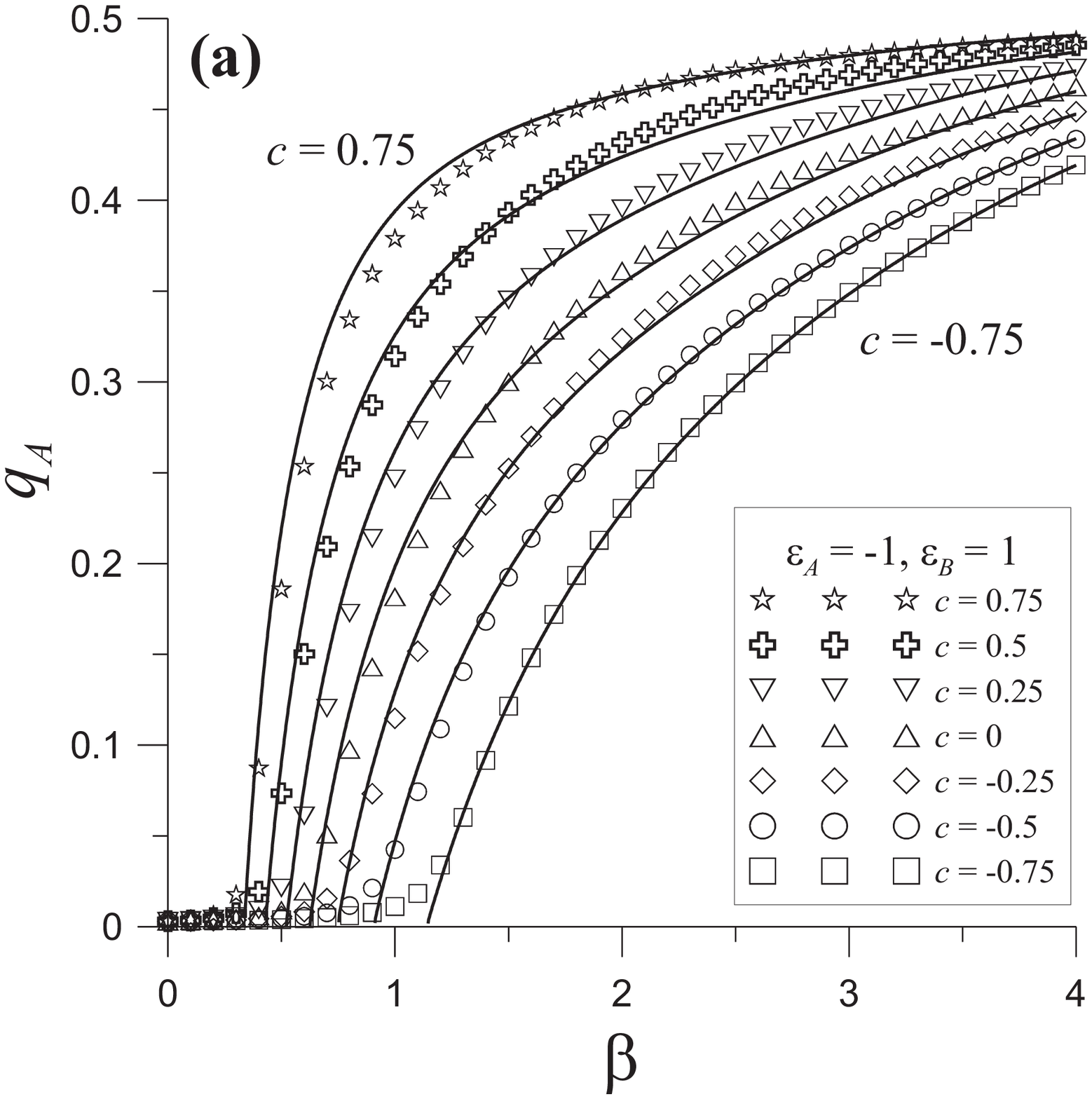}
  \includegraphics[width=7.4cm]{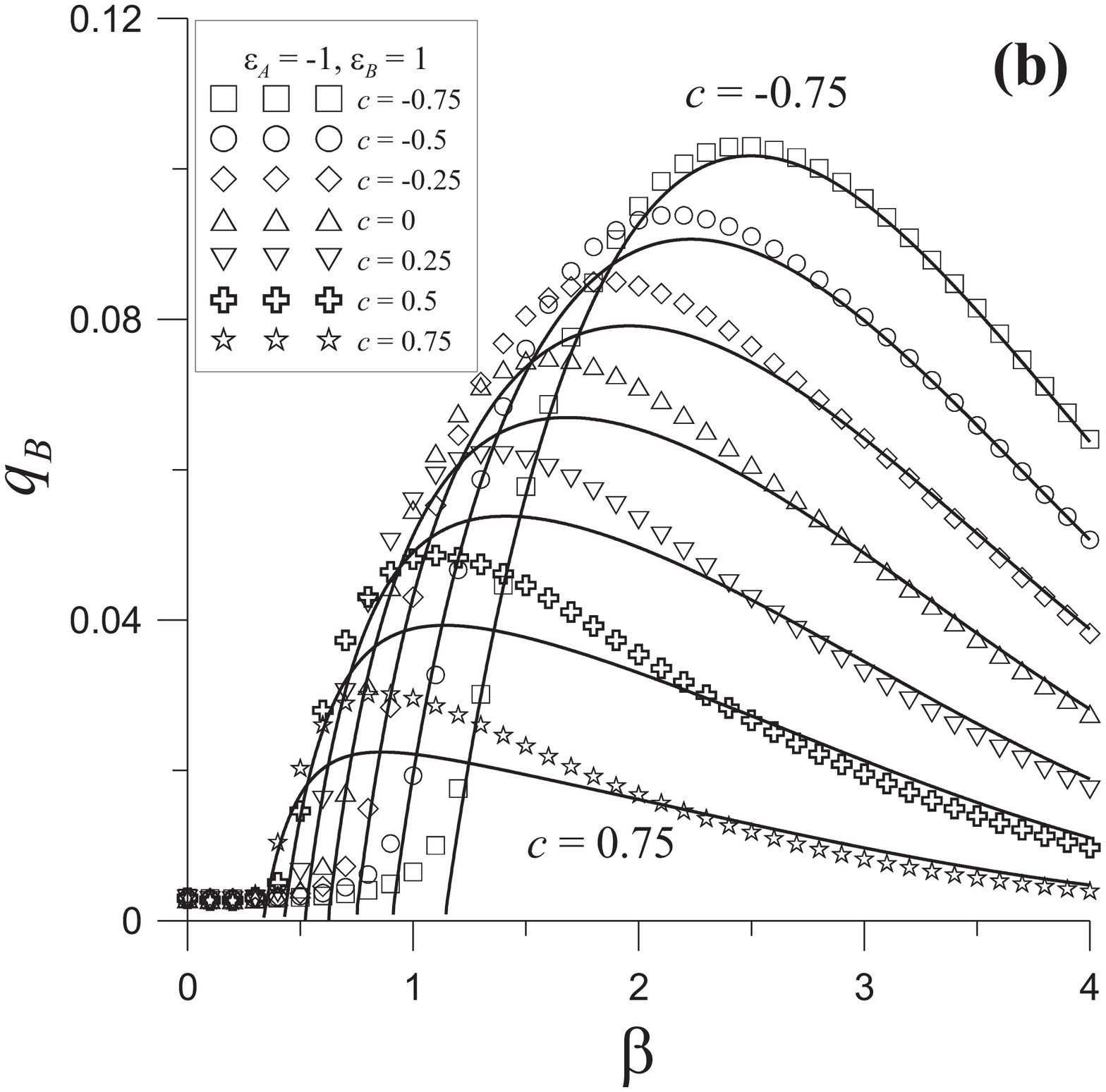}
  \end{center}
  \caption{Fraction of A and B contacts calculated in the Morita approximation (lines) and using the transfer matrix approach
  (symbols) for $f_A=0.5$,  $\varepsilon_A=-1$, $\varepsilon_B=1$ and different values of cluster parameters
  $c=-0.75$~($\Box$), $-0.5$~($\bigcirc$), $-0.25$~($\Diamond$), $0$~($\bigtriangleup$), $0.25$~($\bigtriangledown$),
  $0.5$~($+$), $0.75$~($\star$).}
  \label{fig:Mor:sr}
\end{figure}

Let us turn to the SR case. The results for the monomer-surface
contacts $q_A$ and $q_B$ are presented in Fig.~\ref{fig:Mor:sr}. As
in the SN case, the fraction of A contacts grows with increasing
$\beta$. Furthermore, $q_A$ decreases with decreasing cluster
parameter $c$, in a much more pronounced fashion than for the SN
case. In addition, the growth of $q_A$ with $\beta$ becomes slower
with decreasing $c$.

Unlike $q_A$, the fraction of (unfavourable) B contacts $q_B$
depends non-mo\-no\-to\-ni\-cal\-ly on the inverse temperature,
Fig.~\ref{fig:Mor:sr}~b. It increases just after the transition
point, passes through a maximum and then decays. This behavior looks
similar to that observed in the annealed approximation (both for the
SN and SR cases, Figs.~\ref{fig:ann:sn}~b and \ref{fig:ann:sr}~b),
but its nature is very different: In the annealed case, B-monomers
can convert to A ones, whereas both in the quenched case and in the
Morita approximation a B$\to$A conversion is not possible. The
initial growth of $q_B$ is due to the cooperative effect,
Fig.~\ref{fig:cooperativity}, whereas the subsequent decay is caused
by the increasing influence of the energy penalty on B-contacts.
(Indeed, one can easily imagine the most probable (ground state)
conformation in the $\beta \to \infty$ limit: All A-monomers are on
the surface, all B-monomers are desorbed in loops.) It is worth
noting that the Morita approximation seems to underestimate $q_A$
and overestimate $q_B$ at large $\beta$, {\em i.e.}, small
temperatures. This is counterintuitive, given that the Morita
copolymer can adjust the A-B distribution on the sequence to the
conformation to some extent. In fact, the discrepancy is simply an
effect of the finite chain length in the numerical simulations. On
the one hand, the cooperativity effect is reduced in the free tail,
which reduces the number of B-contacts. On the other hand,
A-monomers close to the tethered end can make contact with the
surface more easily than A-monomers in the middle of the chain,
which enhances the number of A-contacts.

The results presented in Fig.~\ref{fig:Mor:sr} lead us to the
conclusion that the Morita approximation reproduces the fractions of A and B
contacts in the quenched system reasonably well. The quantitative
agreement is best for quasi-alternating RCs ($c=-0.5$ and $-0.75$). We
suggest the following explanation of this effect: The Morita
approximation reduces the initial heteropolymer problem to a
homopolymer problem, thus ``smoothing out'' the monomer sequence of
the RC. Obviously, a quasi-alternating RC can easily (and more
realistically) be represented as a homopolymer with a monomer unit
replaced by an effective one than a RC with $c>0$ where the blocky
structure is essential (it could be better represented by a
multiblock copolymer).

\subsection{Adsorption transition point in the Morita approximation} \label{sec:results:tp}
Solving the problem of RC adsorption in the Morita approximation
amounts to finding the solution of a system of three equations,
Eqs.~(\ref{eq:det:Morita}) and (\ref{eq:MConstr:y}), with three
unknowns: $y_c$, $\lambda$, and $\kappa$ (the partition function/the
smallest singularity of $\Gamma(x)$ is then given by $x_c=y_c
e^{\lambda f_A + \kappa f_{AA}}$).

We obtained that the Lagrange multipliers $\lambda$ abd $\kappa$
vanish in the the adsorption transition point: $\lambda(\beta_{tr})
= \kappa(\beta_{tr}) = 0$. This implies that at the transition
point, the Morita and the annealed approximation are equivalent.
Hence, the transition point is the same for the Morita copolymer
than for the annealed copolymer and is determined by
Eq.~(\ref{eq:tp:annealed}), consistent with the conclusion of
Caravenna and Giacomin~\cite{Caravenna:2005}.
%
\section{Summary and outlook} \label{sec:summary}
In the present work, the adsorption of a single two-letter RC chain with correlations
in its monomer sequence onto a planar homogeneous surface was studied.
One of the monomers (A) was always taken to be attracted to the surface whereas
the other (B) -- was neutral or repulsive (SN: sticky -- neutral or
SR: sticky -- repulsive cases, respectively). The monomer sequence of the RC was
modeled as a first order Markov chain, which is completely determined by
two independent parameters. For the sake of convenience, we used as these
parameters the probability for A monomers in the sequence, $f_A$, and the
cluster parameter $c$ ($c=0$ - uncorrelated (Bernoullian), $c>0$ - quasi-blocky,
and $c<0$ - quasi-alternating copolymer). To average the free energy over the sequence
disorder , {\em i.e.}, over the whole ensemble of monomer sequences, two approximations
were employed: The annealed approximation and the Morita approximation.
The partition function was calculated in the grand canonical ensemble using the
GF approach which is based on the representation of an instantaneous
conformation of the adsorbed chain as a sequence of \emph{independent non-interacting}
trains (adsorbed sequences) and loops flanked by tails.

In the framework of the annealed approximation, where the partition function,
instead of its logarithm, is averaged over sequence disorder, a general approach
for the calculation of the averaged partition function of the RC chain (in the form of
the smallest singularity of the GF), the fraction of adsorbed monomers,
and the transition point was developed. It was found that the annealed approximation
cannot describe the quenched system correctly in the intermediate and strong adsorption
regimes: For example, it strongly overestimates the fraction of contacts of adsorbing
monomers, because of the possibility of A$\to$B monomer conversion. Nevertheless,
the results for annealed chains are interesting in their own right since they
describe the adsorption behavior of so-called two-state polymers (annealed copolymers).

The theory developed for the annealed approximation was then extended and
combined with the Morita approximation. The latter approach is based on a
constrained annealing procedure, where the quenched average is replaced by
an annealed average subject to constraint on the moments of the monomer
distribution. In our case, constraints on the first and the second moments
were imposed.

To estimate the accuracy and validity of the Morita approximation, numerical
calculations using a transfer matrix approach were carried out for the quenched
system: Polymer chains of length $N=1000$ monomers were considered, for the given
monomer statistics 100 realizations of random sequence were generated, for each
realization the free energy and other observables were calculated and then averaged
over the different realizations. The comparison of the results obtained using the
Morita approximation and those for quenched system showed not only a qualitative
but also a good quantitative correspondence, especially for quasi-alternating
RCs ({\em i.e.}, RCs with $c<0$).

It was shown that the fraction of adsorbed A-monomers grows monotonically as
the inverse temperature $\beta$ increases, independently of the character of
the interaction between the nonadsorbing B-monomer and the surface. In contrast,
the fraction of adsorbed B-monomer behaves differently in SN and SR situations:
In the former case, it increases with $\beta$, whereas in the latter case,
it behaves nonmonotonically, exhibiting first a relatively rapid growth just
above the adsorption transition point which is followed by decay at higher values
of $\beta$.

The results of our calculations also indicate that both the annealed and the
Morita approximations give the same adsorption transition point.

The special case $f_A=0.5$, $c=-1$ corresponds to the limiting situation of a
quenched regular alternating AB-copolymer which can be  solved exactly using the
formal framework of the annealed approximation.

We remark that the approach developed in the present work can be applied
in a straightforward way to other types of lattice models of polymers or
to other geometries of adsorbing substrates. The key difference will be the
calculation of GFs for adsorbed sequences
(which is usually quite easy) and loops (which is a more sophisticated task).
In the framework of the GF approach other interesting observables can be
calculated as well, for example, the statistics of trains and loops.
This will be the subject of future work.

\section*{Acknowlegdement}
\label{Acknowlegdement}
The financial support of the Deutsche Forschungsgemeinschaft (SFB 613) is gratefully acknow\-ledged. A.P. thanks Russian Foundation for Basic Research (grant 08-03-00336-a).

\appendix
%
\section{Calculation of $\Gamma(x)$ in the annealed approximation} \label{app:GF:annealed}
Let us tranform the partition function (\ref{eq:GF:annealed}) by exchanging the order of summation over conformations ($\omega$) and disorder ($\chi$) and taking the explicit form of the sequence probability $P(\chi)$, Eq~(\ref{eq:P:def}) and later noting that the inner sum over all realisations of $\chi$ can be written in the matrix form:
\begin{equation} \label{eq:Za:1}
\begin{split}
\left\langle Z_N(\beta|\chi) \right\rangle
  & = \sum_{\chi} P(\chi) \sum_{\omega} \exp\left\{
   -\beta \sum_{i=1}^{N} \Delta_i \cdot \left[ \chi_i \varepsilon_A
   + (1-\chi_i)\varepsilon_B \right] \right\} \\
& = \sum_{\omega} \sum_{\chi} f_{\chi_1} \, e^{-\beta \Delta_1
\cdot \left[ \chi_1 \varepsilon_A  + (1-\chi_1)\varepsilon_B \right]}
\prod_{i=2}^{N} p_{\chi_{i-1}, \, \chi_i} \, e^{-\beta \Delta_i \cdot
\left[ \chi_i \varepsilon_A  + (1-\chi_i)\varepsilon_B \right]}\\
& = \sum_{\omega} (f_A  w_A^{\Delta_1}, \, f_B  w_B^{\Delta_1}) \prod_{i=2}^{N}
\left(
\begin{array}{cc}
p_{AA} w_A^{\Delta_i} & p_{AB} w_B^{\Delta_i} \\
p_{BA} w_A^{\Delta_i} & p_{BB} w_B^{\Delta_i}
\end{array}
\right)
\cdot
\left(
\begin{array}{c}
1 \\
1
\end{array}
\right) \\
& = \sum_{\omega} (\mathbf{W}^{\Delta_1} \mathbf{f})^\mathrm{T} \prod_{i=2}^{N}
(\mathbf{PW}^{\Delta_i}) \cdot \mathbf{e},
\end{split}
\end{equation}
where we have defined $w_A \equiv e^{-\beta\varepsilon_A}$, $w_B \equiv e^{-\beta\varepsilon_B}$, and
\begin{equation} \label{eq:f:Q:def}
\mathbf{f} =
\left(
\begin{array}{c}
f_A  \\
f_B
\end{array}
\right), \quad
\mathbf{W}=
\left(
\begin{array}{cc}
w_A & 0 \\
0 & w_B
\end{array}
\right), \quad
\mathbf{e}=
\left(
\begin{array}{c}
1 \\
1
\end{array}
\right).
\end{equation}

The final result in Eq. (\ref{eq:Za:1}) does not depend on the sequence, {\em i.e.},
we have effectively a homopolymer partition function (but of a somewhat complicated form).
Since $\Delta_i$ can be either 0 or 1, each conformation will give us a product with the
factors $\mathbf{f}^\mathrm{T}$ or $\mathbf{Wf}^\mathrm{T}_1$ (from the first monomer) and $\mathbf{P}$
and $\mathbf{R}\equiv \mathbf{PW}$ (from the monomers with $i=2,3 \ldots N$ ) in a certain order corresponding to this conformation.
Unfortunately, the matrices $\mathbf{P}$ and $\mathbf{R}$ do not commute and
we are not able to simplify this expression by, for example, rewriting the sum over
$\omega$ as a sum over the number of polymer contacts with the surface~\cite{Soteros:2004}.

On the other hand, each adsorbed sequence (train), loop, or tail with the length $n$ contributes the factor $\mathbf{R}^n$, $\mathbf{P}^n$, or $\mathbf{P}^n$, respectively, in the partition function (\ref{eq:Za:1}) (if we do not consider the first monomer) and we can use the sequence GF method (or the grand canonical ensemble technique) suggested by Lifson \cite{Lifson:1964}
Introducing the GF of $\left\langle Z_n(\beta|\chi) \right\rangle$ (or grand canonical p.f.)
$\Gamma(x)=\sum_{n=1}^\infty \left\langle Z_n(\beta|\chi) \right\rangle x^n$
and applying Lifson's arguments~\cite{Lifson:1964} we obtain the result (\ref{eq:GF:annealed}) for $\Gamma(x)$.

\section{Calculation of loop generating function} \label{app:GL}
In this Appendix, we briefly describe the calculation of the loop GF
for the simple (6-choice) cubic lattice. First let us recall the "definition" of the loop.
A loop of length $n$
\begin{itemize}
    \item starts and ends in the closest to the substrate lattice layer (site),
    \item contains \textbf{NO} monomers in the surface layer,
    \item the last, $n$th, monomer is followed by the step in the surface layer,
    \item all the effects (statistical weights) at the \emph{beginning} of a loop
    may be attributed to  the preceding adsorbed segment.
\end{itemize}

\begin{figure}[t]
  \begin{center}
  \includegraphics[width=10cm]{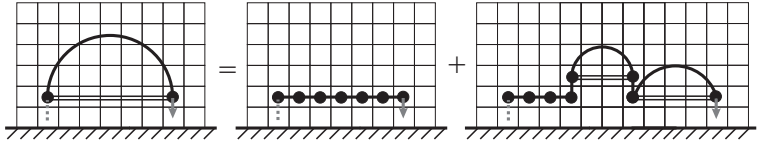}
  \end{center}
  \caption{Decomposition of a loop on 6-choice simple cubic lattice.}
  \label{fig:loop:6scl}
\end{figure}

First, let us notice that the minimum loop length for the given model is equal to 1.
To calculate the GF of a loop
\begin{equation} \label{eq:GL:def}
    \Gamma_L (x) = \sum_{n=1}^\infty \Omega_L(n) x^n
\end{equation}
in the 6-choice simple cubic lattice, we have to take into account two possibilities
(Fig.~\ref{fig:loop:6scl}):
\begin{enumerate}
    \item All the monomers of the loop are in the second layer (First "term" in the rhs of
    Figure \ref{fig:loop:6scl}). The corresponding number of conformations is

    $$(z-2)^{n-1}$$

    \item The first $k$ monomers are in the second layer, then a step into the third layer
    follows. We decompose this contribution in three terms: The first $k$ monomers,
        then loop starting from the second layer, which ends as soon as the first monomer
        reenters the second layer, and the rest.  (Second "term" in the rhs of
        Figure \ref{fig:loop:6scl}). The number of conformations is given by
    $$
      \sum_{k=1}^{n-2} \sum^{l=1}_{n-k-1} (z-2)^{k-1} \, \Omega_L(l) \, \Omega_L(n-k-l) \, ,
      \quad n \geq 3
    $$
\end{enumerate}
This gives us a recursion relation for $\Omega_L(n)$:
\begin{equation}
      \Omega_L(n) = (z-2)^{n-1} + \sum_{k=1}^{n-2} \sum^{l=1}_{n-k-1} (z-2)^{k-1} \,
      \Omega_L(l) \, \Omega_L(n-k-l) \, , \quad n \geq 3 \, .
\end{equation}
By multiplying this equation by $x^n$ and summing from $n=3$ to infinity,
we obtain the following recursion relation for the GF
$\Gamma_L(x)$ (\ref{eq:GL:def})
\begin{equation}
    \Gamma_L(x) - x - (z-2)x^2 = \frac{x^3 (z-2)^2}{1-(z-2) x} +
    \frac{x}{1-(z-2) x}\Gamma_L^2(x) \, .
\end{equation}
Its solution is
\begin{equation}
  \Gamma_L(x)=\frac{1}{2x}\left\{ 1- (z-2)x - \sqrt{(1-zx)[1-(z-4)x]} \right\} \, .
\end{equation}
%
%

%

\end{document}